\documentclass[aps,pra,amsmath,amsfonts,amssymb,twocolumn,nofootinbib]{revtex4}

\usepackage{amssymb}
\usepackage{graphicx,psfrag,color}
\usepackage{dcolumn}
\usepackage{bm}
\usepackage{mathbbol}
\usepackage[T1]{fontenc}

\newcommand{\med}[1]{\left\langle #1 \right\rangle}

\begin{document}
\flushbottom

\title{Finite temperature detection of quantum critical points via internal 
quantum teleportation}

\author{G. A. P. Ribeiro}
\author{Gustavo Rigolin}
\email{rigolin@ufscar.br}
\affiliation{Departamento de F\'isica, Universidade Federal de
S\~ao Carlos, 13565-905, S\~ao Carlos, SP, Brazil}

\date{\today}

\begin{abstract}
We show that the teleportation protocol can be efficiently used to 
detect quantum critical points using finite temperature data even if all
resources needed for its implementation lie within the system under 
investigation. 
Contrary to a previous proposal, there is no need to use an external 
qubit as the input state to be teleported 
to one of the qubits within the system. Here,
we use a pair of nearest neighbor spins 
from an infinite spin-1/2 chain in equilibrium with a heat bath as the 
entangled resource of the quantum teleportation protocol and a third 
adjacent qubit within the chain itself as the input state to be teleported. 
For several spin chain models subjected to an external magnetic 
field, we show that the efficiency of the teleportation protocol
is severely affected as we cross the quantum critical points associated with
those spin chains. This abrupt change in efficiency gives us a clear indication of a quantum phase transition.
\end{abstract}


\maketitle

\section{Introduction} 

The qualitative change in the macroscopic properties of a many-body system 
that occurs at the absolute zero temperature ($T=0$) as we change its
Hamiltonian is called a quantum phase transition (QPT).
Theoretically it is described by the change of the many-body system's 
ground state as we slowly (adiabatically) modify its Hamiltonian ($H$) \cite{sac99,gre02,gan10,row10}. Since at $T=0$ there are no thermal fluctuations,
this change in the physical properties of the many-body system is caused 
by quantum fluctuations whose origin can be traced back to the 
Heisenberg uncertainty principle. A standard example of a
QPT is the onset of a non-null magnetization when a spin chain 
enters its ferromagnetic phase coming from the paramagnetic one. In 
general, a QPT is given by a symmetry change in the 
ground state of the system and the appearance of an order parameter 
(the non-null magnetization in the above example).

Many important quantum information theory quantities were shown to be 
excellent quantum critical point (QCP) detectors at $T=0$
\cite{ost02,osb02,wu04,oli06,dil08,sar08,fan14,gir14,hot08,hot09,
ike23a,ike23b,ike23c,ike23d}. However, 
it is known that some of them do not work properly when $T>0$. 
For example, the entanglement 
of formation \cite{woo98} between a pair of spins that belongs to a spin
chain is zero before, at, and after the QCP if the chain
is prepared above a certain threshold temperature \cite{wer10,wer10b}. 
Therefore,
both from a theoretical point of view and from practical aspects
(we cannot cool a many-body system to $T=0$ due to the third law of thermodynamics), it is important to develop robust tools to characterize 
QPTs at finite $T$. In this case thermal
fluctuations are present and we must have tools to detect QCPs that 
properly work in this scenario.

One of the most successful and resilient tool to pinpoint a QCP at finite 
$T$ is quantum discord (QD) \cite{wer10b}, originally defined in Refs. \cite{oll01,hen01}. From a theoretical point of view, though, the 
computation of QD is not feasible for high spin systems. 
The evaluation of QD is an NP-complete problem \cite{hua14}, which means that 
the evaluation of QD becomes intractable as we increase the system's Hilbert space
dimension. Even for spin-1 systems one already faces great challenges to compute it \cite{mal16}. 
Moreover, from an experimental point of view, QD has no operational meaning.
So far there is no general experimental procedure to directly measure QD. 
The determination of QD for a given system can only be achieved after 
we obtain either theoretically or experimentally its complete density matrix.

In Ref. \cite{pav23} we went one step further and presented QCP detection 
tools that keep the most useful features of QD in detecting QCPs 
at finite $T$ but do not have its theoretical and experimental handicaps 
outlined above. These tools are based on the quantum teleportation 
protocol \cite{ben93,yeo02,rig17} and are scalable to high dimensional
systems, have a clear experimental meaning, and work at temperatures where
other tools, such as the entanglement of formation, already fail to spotlight QCPs. See also Refs. \cite{ven06,ven07,ven07b,aba23} 
for other applications of the quantum teleportation protocol in  
spin chains, in particular the teleportation of a quantum 
state between distant spins of the chain.

In this work we show, under very general conditions, that we can
simplify the tools to detect QCPs at finite $T$ given in Ref. \cite{pav23},
while still keeping their efficiency, robustness, and scalability. 
In the present proposal, we do not need an external qubit from the 
system to implement the teleportation based QCP detector. We do
not even need
to repeat the procedure using several different external qubits 
covering the whole Bloch sphere in order to detect a QCP \cite{pav23}. 
In the present proposal, the input state to be teleported is now fixed and  belongs to the many-body system itself (see Fig. \ref{fig_scheme}).  
Also, and similarly to the tool developed in Ref. \cite{pav23}, we do not 
need to know the order parameter associated with the QPT to implement it
(see Refs. \cite{wer10b,pav23} for more details).

Being more specific, we study the ability of the present tools, 
developed in Sec. \ref{tool}, to detect QCPs at finite or zero temperature 
for several spin-1/2 chains in the thermodynamic limit 
(infinite chains). We study the XXZ model subjected to a longitudinal external magnetic field in Sec. \ref{secXXZ} and the Ising model and the XY model in a transverse magnetic field in Sec. \ref{secXY}. We end this work by giving 
in Sec. \ref{discussion} an extensive analysis of the theoreti\-cal and experimental resources needed
to apply these tools and in Sec. \ref{conclusion} we provide 
our concluding remarks.

\section{The critical point detector}
\label{tool}

\subsection{General settings}
\label{gs}

\begin{figure}[!ht]
\includegraphics[width=6cm]{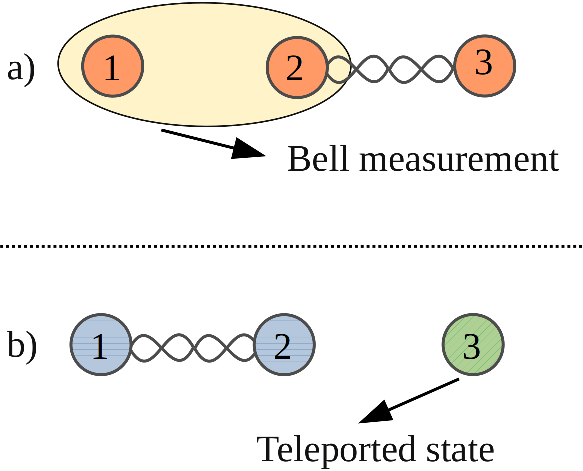}
\caption{\label{fig_scheme}(color online) 
One run of the teleportation protocol can be essentially divided into 
four steps, condensed in the two panels above. 
First, Alice and Bob share an entangled resource
(qubits $2$ and $3$) and Alice choose a qubit to be teleported (qubit
$1$). Second, Alice implements a Bell measurement (BM) 
on qubits $1$ and $2$. A BM is a joint 
measurement on the qubit to be teleported to Bob and on one of
the qubits of the entangled state shared with Bob. This step of the protocol is
depicted in panel (a) above. Third, Alice informs Bob of her measurement 
result. There are four possibilities here and she needs to send two classical
bits to Bob in order to tell him about which Bell state she measured.
Fourth, Bob applies an appropriate unitary operation on his qubit depending on
which Bell state Alice measured. After this step, the protocol ends and it is
illustrated in panel (b). Bob's qubit at the end of a run of the protocol 
is more or less similar to Alice's original input state according to 
the quality of the shared entangled state between them (qubits $2$ and $3$).
}
\end{figure}

The key ingredient in our QCP detector is the standard teleportation protocol \cite{ben93}, expressed in the mathematical language of density matrices 
instead of pure states \cite{pav23,rig17,rig15}. The 
entangled resource shared by Alice and Bob is given by qubits $2$ and $3$,
respectively, and are described by the density matrix $\rho_{23}$. 
Alice's input qubit, the one to be teleported to Bob, is given by the 
density matrix $\rho_1$. See Fig.~\ref{fig_scheme} for a step by step
description of the protocol.

Initially, the total state describing the three qubits participating in
the teleportation protocol is
\begin{equation}
 \rho = \rho_{1} \otimes \rho_{23}
\label{stepA}
\end{equation}
and after one run of the protocol Bob's qubit is 
\cite{pav23,rig17}
\begin{equation}
\rho_{_{B_j}}=   \frac{U_jTr_{12}[P_j \rho P_j]U_j^\dagger}{Q_j}.
\label{stepD}
\end{equation}
In Eq.~(\ref{stepD}), $Tr_{12}$ is the partial trace over Alice's qubits 
(spins 1 and 2), $j$ is the Bell measurement (BM) outcome
obtained by Alice ($j=\Psi^{\pm},\Phi^{\pm}$), and $P_j$ 
denotes the four projectors associated with the BMs,
\begin{eqnarray}
P_{\Psi^{\pm}} &=& |\Psi^{\pm}\rangle \langle \Psi^{\pm}|, \label{projectorA}\\  
P_{\Phi^{\pm}} &=& |\Phi^{\pm}\rangle \langle \Phi^{\pm}|, \label{projectorB}  
\end{eqnarray}
where
\begin{eqnarray}
|\Psi^{\pm}\rangle&=&(|01\rangle \pm |10\rangle)/\sqrt{2},\label{BellA} \\ 
|\Phi^{\pm}\rangle&=&(|00\rangle \pm |11\rangle)/\sqrt{2}. \label{BellB}
\end{eqnarray}
The states given by Eqs.~(\ref{BellA})-(\ref{BellB})
are the four Bell states.

The other two quantities appearing in Eq.~(\ref{stepD}) are 
Alice's probability to obtain a certain Bell state $j$ \cite{pav23,rig17},
\begin{equation}
Q_j = Tr[{P_j \rho}],
\label{prob}
\end{equation}
and the unitary correction $U_j$ that Bob must apply on his qubit after being informed by Alice of her measurement outcome. 

The unitary correction $U_j$ that Bob must implement on his qubit after a  
run of the protocol depends not only on Alice's measurement result but also
on the entangled state $\rho_{23}$ shared with Alice.  
The sets $S_k$ below give the
unitary operations that Bob must apply on his qubit if the shared state between Alice and Bob is the Bell state $|k\rangle$ \cite{pav23,rig17},
\begin{eqnarray}
S_{\Phi^+}=\{U_{\Phi^+},U_{\Phi^-},U_{\Psi^+},U_{\Psi^-}\}
=\{\mathbb{1},\sigma^z,\sigma^x,\sigma^z\sigma^x\},
\label{s1} \\
S_{\Phi^-}=\{U_{\Phi^+},U_{\Phi^-},U_{\Psi^+},U_{\Psi^-}\}
=\{\sigma^z,\mathbb{1},\sigma^z\sigma^x,\sigma^x\}, \label{s2}\\
S_{\Psi^+}=\{U_{\Phi^+},U_{\Phi^-},U_{\Psi^+},U_{\Psi^-}\}=\{\sigma^x,\sigma^z\sigma^x,\mathbb{1},\sigma^z\}, \label{s3}\\  
S_{\Psi^-}=\{U_{\Phi^+},U_{\Phi^-},U_{\Psi^+},U_{\Psi^-}\}=\{\sigma^z\sigma^x,\sigma^x,\sigma^z,\mathbb{1}\}.\label{s4}
\end{eqnarray}
Note that $\mathbb{1}$ is the identity matrix and $\sigma^\alpha$, 
with $\alpha=x,y,z$, is the standard Pauli matrix \cite{nie00}. 

The state $\rho_{23}$ shared 
between Alice and Bob depends on the system's quantum phase. 
In one quantum phase $\rho_{23}$ 
is more similar to one of the four Bell states while in a different quantum
phase it is best approximated by another one. Hence, in order to better
characterize the QCPs of a spin chain we will use the four sets $S_k$
of unitary operations and also determine the set that gives the 
most efficient teleportation protocol within a quantum phase.  

The efficiency of the teleportation protocol is given by how close 
the output state $\rho_{_{B_j}}$ is to the input state $\rho_1$ 
at the end of a run of the protocol. Therefore, in order to quantitatively 
assess that efficiency, we need to choose quantitative measures of the similarity between the teleported state at the end of the protocol and the initial input state with Alice. 

In this work we will be dealing with two measures that quantify the 
similarity between those two states. The first one is the fidelity \cite{uhl76}, which is given by \cite{nie00}  
\begin{equation}
F_j(S_k)=F(\rho_1,\rho_{_{B_j}}) =  
\left[Tr\sqrt{\sqrt{\rho_1}\rho_{_{B_j}}\sqrt{\rho_1}}\right]^2.
\label{Fidj}
\end{equation}
In the left hand side of Eq.~(\ref{Fidj}) we made it explicit that 
the fidelity also depends on the set $S_k$ of unitary operations 
that Bob must apply on his qubit. If the teleported state is 
exactly equal to Alice's input state, we obtain $F_j=1$, and 
$F_j=0$ if those states are orthogonal. Note that in Ref. \cite{nie00}
the fidelity is defined as
$Tr\sqrt{\sqrt{\rho_1}\rho_{_{B_j}}\sqrt{\rho_1}}$. Here we take 
the square of the previous expression to conform with the definition used
in our previous works when the input state was a pure state \cite{pav23} and with the original work of Uhlmann \cite{uhl76}.

After several implementations (runs) of the protocol, the mean fidelity (or efficiency of 
the protocol) is \cite{pav23,rig17,rig15,gor06}
\begin{equation}
\overline{F}(S_k)= \sum_{j=\Psi^{\mp},\Phi^{\mp}} 
Q_jF_j(S_k). \label{Fbar}
\end{equation}
Similarly to Ref. \cite{pav23}, Eq.~(\ref{Fbar}) is the basic building block
from which one teleportation based QCP detector is built. In particular,
it is defined as the maximum mean fidelity \cite{pav23}
\begin{equation}
\overline{\mathcal{F}} = \max_{\{S_k\}}{\overline{F}(S_k)}.
\label{fmax}
\end{equation}
As we will show in the next sections, the functional behavior of 
$\overline{\mathcal{F}}$ and of its first and second
order derivatives in terms of the tuning parameter that drives the change in the
system's Hamiltonian is a very useful QCP detector at zero and finite temperatures.

The second measure we will be using in this work to quantify the similarity 
between two mixed states is the trace distance \cite{nie00,tos23,tos24},
\begin{equation}
D_j(S_k)=D(\rho_1,\rho_{_{B_j}}) = \frac{1}{2}Tr
\left|\rho_1 - \rho_{_{B_j}}\right|,
\label{Dj}
\end{equation}
where for an operator $A$ we have $|A| = \sqrt{A^\dagger A}$.  
A direct computation for spin-1/2 density matrices gives
\begin{equation}
D_j(S_k) = \frac{1}{2}\sqrt{(\Delta r_x)^2+(\Delta r_y)^2+(\Delta r_z)^2},
\label{Dj2}
\end{equation}
where $\Delta r_\alpha(t) = Tr(\rho_1\sigma^\alpha)-Tr(\rho_{_{B_j}}\sigma^\alpha)$,
with $\sigma^\alpha$ being a Pauli matrix. Geometrically, the trace 
distance is half the Euclidean distance between the points on the Bloch sphere 
associated with the density matrices $\rho_1$ and $\rho_{_{B_j}}$ \cite{nie00}. 
Therefore, when the two states are equal we have $D_j=0$ and the farther apart
the two states are in the Bloch sphere the greater $D_j$ is. For orthogonal pure 
states we have $D_j=1$, the maximum possible value for the trace distance.

After several runs of the teleportation protocol, the mean trace distance is given
by
\begin{equation}
\overline{D}(S_k)= \sum_{j=\Psi^{\mp},\Phi^{\mp}} 
Q_jD_j(S_k) \label{Dbar}
\end{equation}
and the second QCP point detector is built by minimizing 
$\overline{D}(S_k)$ 
over all sets $S_k$,
\begin{equation}
\overline{\mathcal{D}} = \min_{\{S_k\}}{\overline{D}(S_k)}.
\label{dmin}
\end{equation}

Similarly to $\overline{\mathcal{F}}$, the behavior of 
$\overline{\mathcal{D}}$ and of its first and second
order derivatives with respect to the Hamiltonian's  
driving parameter turns out to be
a very robust QCP detector at zero and finite temperatures.
 
\textit{Remark 1:} Note that a high value for the fidelity between two states 
implies a low value for the trace distance and vice-versa. The fidelity measures
the similarity between two states; the more similar the states the greater the fidelity. 
The trace distance, as its name suggests, measures the distance between those two
states; the closer the states the lower the value of the trace distance. This is 
why for the fidelity we implement a maximization operation while for the 
trace distance we implement a minimization one.

\textit{Remark 2:} The reason we use in this work the trace distance in addition to
the fidelity to assess the efficiency of the teleportation protocol is related to
two facts. First, the expressions for the fidelity between two mixed states are 
too cumbersome and complicated, while the expressions for the trace distance are 
relatively simpler. Second, the fidelity is less efficient than the trace distance
to detect one of the QCPs of the XXZ model (see Sec. \ref{secXXZ}). For the other
models and the other QCP of the XXZ model, both quantities are for all
practical purposes equivalent.

\subsection{Specific settings}
\label{ss}

All the models we study here (Secs. \ref{secXXZ} and \ref{secXY}) are one dimensional spin-1/2 chains in the thermodynamic
limit ($L\rightarrow \infty$) satisfying periodic boundary conditions
($\sigma^{\alpha}_{L+1} = \sigma^{\alpha}_1$). Here the subscript 
implies that $\sigma^{\alpha}_j$ acts on the qubit located at the lattice site $j$.
Also, all spin chains are assumed to be in equilibrium with a heat bath (thermal reservoir at temperature $T$) and, as such, the density matrix describing the whole chain is given by the canonical ensemble density matrix 
\begin{equation}
\varrho=e^{-H/kT}/Z, \label{canonical} 
\end{equation}
with $Z=Tr[e^{-H/kT}]$ being the partition function and $k$ the Boltzmann's constant. 

The density matrix describing three nearest
neighbors from the spin chain is given by $\rho_{123}$, which
is obtained by tracing out from $\varrho$ all but spins 
$1,2$ and $3$. Note that the initial state $\rho_{123}$ can be obtained using any three nearest neighbors from the spin chain. After putting the whole chain
in contact with a heat reservoir at temperature $T$ and waiting 
for the system to reach the thermodynamic equilibrium, any three 
consecutive spins
of the chain can be used to implement the internal teleportation protocol as described below.

The density matrix $\rho_{123}$ is in general not given by
Eq.~(\ref{stepA}), where $\rho_1$ is given by tracing out from 
$\varrho$ all but spin $1$ and $\rho_{23}$ is the state obtained
by tracing out from $\varrho$ all but spins $2$ and $3$. 
The state $\rho_1$ describes a single spin from the chain 
and $\rho_{23}$ any nearest neighbor pair of spins from the chain. We have,
therefore, to prepare an ensemble from $\rho_{123}$ that is 
effectively given by Eq.~(\ref{stepA}). Before we explain how
this can be done, we first give the explicit forms of  
$\rho_{23}$ and $\rho_1$ in the standard basis.

For all the models investigated here we have \cite{wer10b,pav23,osb02}
\begin{equation}
\rho_{23}  = \left(
\begin{array}{cccc}
a & 0 & 0 & e\\
0 & b  & c & 0 \\
0 & c & b & 0 \\
e & 0 & 0 & d\\
\end{array}
\right), \label{rhoAB2}
\end{equation}
where
\begin{eqnarray}
a &=& \frac{1+2\med{\sigma^z_2}+\med{\sigma_2^z\sigma_{3}^z}}{4}, \\
b &=& \frac{1-\med{\sigma_2^z\sigma_{3}^z}}{4}, \\
c &=& \frac{\med{\sigma_2^x\sigma_{3}^x}+\med{\sigma_2^y\sigma_{3}^y}}{4}, \\
d &=&  \frac{1-2\med{\sigma^z_2}+\med{\sigma_2^z\sigma_{3}^z}}{4}, \\
e &=& \frac{\med{\sigma_2^x\sigma_{3}^x}-\med{\sigma_2^y\sigma_{3}^y}}{4}.
\end{eqnarray}
For the XXZ and XX models $\med{\sigma_2^x\sigma_{3}^x}=\med{\sigma_2^y\sigma_{3}^y}$ while for the Ising and XY models $\med{\sigma_2^x\sigma_{3}^x}\neq \med{\sigma_2^y\sigma_{3}^y}$. We should also note that the translational 
invariance
of the spin chain implies that $\med{\sigma^\alpha_j}=
\med{\sigma^\alpha_k}$ and 
$\med{\sigma^\alpha_j\sigma^\beta_{j+1}}=
\med{\sigma^\alpha_k\sigma^\beta_{k+1}}$, for any value of $j,k$.

The evaluation in the thermodynamic limit and 
for arbitrary values of $T$ of the one-point and two-point correlation functions,
\begin{eqnarray}
z&=&\med{\sigma_j^z}=Tr[\sigma_j^z\, \varrho], \label{z} \\
\alpha\alpha&=&\med{\sigma_j^\alpha\sigma_{j+1}^\alpha}=
Tr[\sigma_j^\alpha\sigma_{j+1}^\alpha\, \varrho], \label{zz}
\end{eqnarray}
%
where $\alpha=x,y,z$, can be found in Refs. \cite{yan66,clo66,klu92,bor05,boo06,boo08,tri10,tak99,lie61,bar70,bar71,pfe70,zho10}. 
In Ref. \cite{wer10b} these calculations are reviewed and 
written in the present notation. The behavior of $\med{\sigma_j^z}$ and
$\med{\sigma_j^\alpha\sigma_{j+1}^\alpha}$ for several values
of $T$ as we change the tuning parameter for each one of the models 
given in Secs. \ref{secXXZ} and \ref{secXY} are shown in Ref. \cite{pav23}.

The density matrix $\rho_1$ describing a single spin 
is calculated by tracing out from 
the thermal state $\varrho$ [Eq.~(\ref{canonical})] all but one spin. Or, equivalently, by tracing out
one of the spins from $\rho_{23}$. Therefore, a simple calculation gives
\begin{equation}
\rho_1  = \left(
\begin{array}{cc}
p_0 & 0\\
0 & p_1 
\end{array}
\right) = p_0|0\rangle\langle 0 | + p_1|1\rangle\langle 1 |, \label{rho1}
\end{equation}
where 
\begin{eqnarray}
p_0 &=& \frac{1 + \med{\sigma^z}}{2}, \label{p0}\\
p_1 &=& \frac{1 - \med{\sigma^z}}{2}. \label{p1}
\end{eqnarray}
In Eqs.~(\ref{p0}) and (\ref{p1}) 
we have dropped the subscript from the one-point correlation function for simplicity.

Using Eq.~(\ref{rho1}) and that 
$$\rho_{23}=Tr_1[\rho_{123}]= \,_1\langle 0| \rho_{123}|0\rangle_1
+\,_1\langle 1| \rho_{123}|1\rangle_1,$$
the state (\ref{stepA}) can be written as follows,
\begin{eqnarray}
\rho &=& p_0|0\rangle_1\,_1\langle 0 | \otimes 
\,_1\langle 0| \rho_{123}|0\rangle_1 \nonumber \\
&&+p_0|0\rangle_1\,_1\langle 0 | \otimes 
\,_1\langle 1| \rho_{123}|1\rangle_1 \nonumber \\
&&+p_1|1\rangle_1\,_1\langle 1 | \otimes 
\,_1\langle 0| \rho_{123}|0\rangle_1 \nonumber \\
&&+p_1|1\rangle_1\,_1\langle 1 | \otimes 
\,_1\langle 1| \rho_{123}|1\rangle_1. \label{rhoalt}
\end{eqnarray}

Writing $\rho$ as given by Eq.~(\ref{rhoalt}) readily suggests
the steps Alice needs to do to effectively obtain
state (\ref{stepA}). She projects qubit 1 in the
computational basis. She obtains either $|0\rangle_1$
or $|1\rangle_1$ with probabilities $p_0$ and $p_1$, 
respectively. According to the measurement 
postulate of quantum mechanics \cite{nie00}, 
the state $\rho_{123}$
changes in each case to 
\begin{eqnarray}
\rho_{123} \rightarrow 
\frac{P_{|0\rangle} \rho_{123} P_{|0\rangle}}{p_0} 
= |0\rangle_1\,_1\langle 0 | \otimes 
\frac{\,_1\langle 0| \rho_{123}|0\rangle_1}{p_0},
\label{ensemble1}
\\
\rho_{123} \rightarrow 
\frac{P_{|1\rangle} \rho_{123} P_{|1\rangle}}{p_1} 
= |1\rangle_1\,_1\langle 1 | \otimes 
\frac{\,_1\langle 1| \rho_{123}|1\rangle_1}{p_1},
\label{ensemble2}
\end{eqnarray}
where $P_{|j\rangle}=|j\rangle_1\,_1\langle j|$, with
$j=0,1$, is the projector onto the state $|j\rangle_1$.

Also, if after measuring her qubit Alice applies the spin
flip unitary operation, namely, 
$\sigma_1^x|0(1)\rangle=|1(0)\rangle$, we have for each 
possible measurement result,
\begin{eqnarray}
\rho_{123} \rightarrow 
\frac{\sigma_1^xP_{|0\rangle} \rho_{123} 
P_{|0\rangle}\sigma_1^x}{p_0} 
= |1\rangle_1\,_1\langle 1 | \otimes 
\frac{\,_1\langle 0| \rho_{123}|0\rangle_1}{p_0},
\label{ensemble3}
\\
\rho_{123} \rightarrow 
\frac{\sigma_1^xP_{|1\rangle} \rho_{123} 
P_{|1\rangle}\sigma_1^x}{p_1} 
= |0\rangle_1\,_1\langle 0 | \otimes 
\frac{\,_1\langle 1| \rho_{123}|1\rangle_1}{p_1}.
\label{ensemble4}
\end{eqnarray}

From four subensembles, each one with 
an equal number $N$ of systems described by Eqs.~(\ref{ensemble1}), (\ref{ensemble2}), (\ref{ensemble3}),
or (\ref{ensemble4}), Alice builds the following ensemble. She picks $(p_0)^2N$ 
states given by (\ref{ensemble1}), $(p_1)^2N$ 
states given by (\ref{ensemble2}), $p_0p_1N$ 
states given by (\ref{ensemble3}), and $p_0p_1N$ 
states given by (\ref{ensemble4}). Combining these four
subensembles leads to $N$ states described by Eq.~(\ref{rhoalt}),
or equivalently, Eq.~(\ref{stepA}).\footnote{
It is worth
noticing that if Alice picks qubit $1$ sufficiently away from 
qubits $2$ and $3$, the state $\rho_{123}$ is already given by
Eq.~(\ref{stepA}). Also, if we prepare two identical spin chains
and pick qubit $1$ from one chain and a pair of nearest neighbors
from the other chain, we also have Eq.~(\ref{stepA}) describing
those three qubits.} 

Note that due to the linearity of all the steps related to the
implementation of the teleportation protocol, from the
experimental point of view Alice's job to effectively obtain Eq.~(\ref{stepA}) is very simple. She just needs to measure 
the qubit $1$ in the computational basis and then either 
do nothing or implement the bit flip operation before initiating the standard teleportation protocol as described in Sec. \ref{gs}. 
Being more specific, if she measures $|0\rangle$ she 
does nothing to her qubit $(p_0)^2$ of the times  
or she flips it $p_0p_1$ of the times before starting the 
teleportation protocol. If she measures $|1\rangle$, she 
does nothing $(p_1)^2$ of the times or she flips it 
$p_0p_1$ of the times before implementing the teleportation
protocol. Also, she must work with an equal number of 
cases where she measured $|0\rangle$ or $|1\rangle$ 
to mimic the state given by Eq.~(\ref{rhoalt}) following the 
prescription given above. This is 
easily accomplished discarding the exceeding cases.

\begin{figure}[!ht]
\includegraphics[width=8cm]{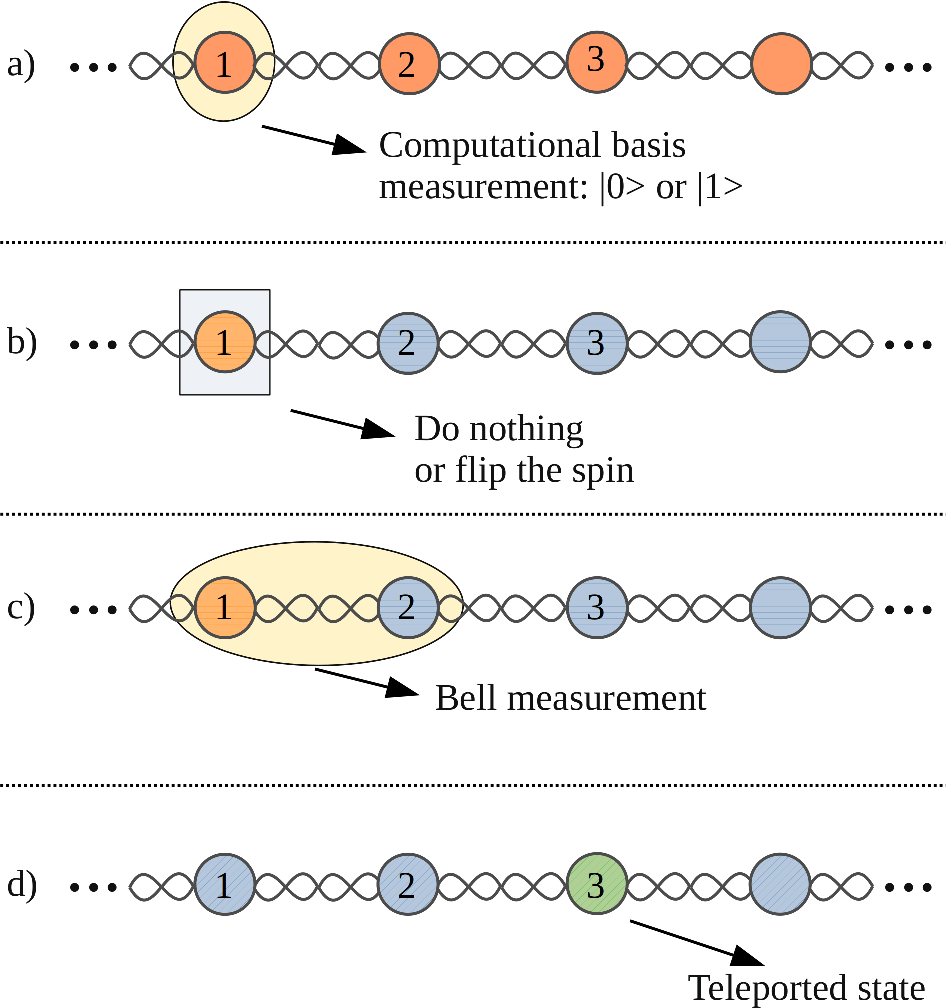}
\caption{\label{fig_scheme2}(color online) 
One run of the internal teleportation protocol can be 
described as follows. 
First, Alice projects the qubit to be teleported (qubit
$1$) onto the computational basis [panel (a)]. 
Second, she either does nothing with her qubit or applies 
onto it the spin flip operation ($\sigma_1^x$) according to the prescription given in the text [panel (b)]. Third, Alice and 
Bob execute the standard teleportation protocol following the steps given in Fig. \ref{fig_scheme} [panels (c) and (d)].
}
\end{figure}

The bottom line is that by proceeding as described above
and due to the linearity of all the 
steps involved in the execution
of the teleportation protocol, the effective initial state will be given by Eq.~(\ref{stepA}) and Bob's qubit at the end of several runs of the teleportation protocol will effectively be described by Eq.~(\ref{stepD}) whenever Alice measures the Bell state $j$.
Figure \ref{fig_scheme2} illustrates all the steps of a 
single run of the protocol.

Alternatively one could use the 
state $\rho_{123}$ to theoretically 
simulate the execution of the internal teleportation protocol.
The state $\rho_{123}$ is a $8\times 8$
matrix depending not only on one- and two-point correlation functions but also on the genuine three-point ones 
($\langle\sigma^\alpha_j\sigma^\beta_{j+1}\sigma^\gamma_{j+1}\rangle$). 
Nevertheless, due to the factorization property of the correlation functions of the XXZ spin chain \cite{boo06}, the three-point correlations for this model can be expressed 
very non-trivially in terms of the one- and two-point correlations.
However, 
the computation of the three-point correlation functions 
for finite temperature and for the other models 
here investigated is a
quite involved task \cite{boo06,boo08,tri10}.
Thus, in order to test the 
usefulness of the internal teleportation QCP detector, 
we worked with an initial state that depends at most on two-point 
correlation functions 
of two consecutive spin. 
It would be very interesting to extend the present analysis 
to the case where the state $\rho_{123}$ is used as the 
initial state. We believe that the internal teleportation 
QCP detector will also work in this case. Perhaps giving results
even better than the ones reported here. We should also note that
from the experimental point of view working with $\rho_{123}$
is even simpler. In this case Alice will not need to project qubit $1$ onto the computational basis or flip her spin before 
implementing the standard teleportation protocol. 


We now return to the computation of the relevant quantities needed
to assess the efficiency of the teleportation protocol. 
Inserting Eqs.~(\ref{stepA}), (\ref{projectorA}), (\ref{projectorB}), (\ref{rhoAB2}), and (\ref{rho1}) into Eq.~(\ref{prob}) gives
\begin{eqnarray}
Q_{\Psi^{\pm}} &=& (1-z^2)/4, \label{qpsi}\\
Q_{\Phi^{\pm}} &=& (1+z^2)/4, \label{qphi}
\end{eqnarray}
where $z$ is given by Eq.~(\ref{z}).

Using now Eqs.~(\ref{Fidj}), (\ref{qpsi}) and (\ref{qphi}), the mean fidelity given by Eq.~(\ref{Fbar}) becomes, 
\begin{widetext}
\begin{eqnarray}
\overline{F}(S_{\Psi^{\pm}}) 
&=& \frac{1}{8}\left(\sqrt{(1 + z) f(zz, -z)}+\sqrt{(1 - z) f(zz, z)}\right)^2
+\frac{1}{8}\left(\sqrt{(1 - z) g(zz, z)}+\sqrt{(1 + z) g(zz, -z)}\right)^2, \label{f1} \\ 
\overline{F}(S_{\Phi^{\pm}}) 
&=& \frac{1}{8}\left(\sqrt{(1 - z) f(zz, -z)}+\sqrt{(1 + z) f(zz, z)}\right)^2
+\frac{1}{8}\left(\sqrt{(1 + z) g(zz, z)}+\sqrt{(1 - z) g(zz, -z)}\right)^2, \label{f4}
\end{eqnarray}
\end{widetext}
where
\begin{eqnarray}
f(zz,z)&=& 1 + z + z^2 + z \cdot zz, 
\label{f}\\
g(zz,z)&=& 1 - z - z^2 + z \cdot zz. \label{g}
\end{eqnarray}
Note that the dot between $z$ and $zz$ in Eqs.~(\ref{f}) and (\ref{g})
means the standard multiplication between the one-point correlation function
$z$ and the two-point correlation function $zz$, Eqs.~(\ref{z}) and (\ref{zz})
respectively.

As anticipated in the second remark at the end of Sec. \ref{gs},
the expressions for the fidelities are not simple and the maximum mean fidelity 
becomes
\begin{equation}
\overline{\mathcal{F}} = \max{\left[\,\overline{F}(S_{\Psi^{\pm}}), 
\overline{F}(S_{\Phi^{\pm}})\right]},
\label{fmax2}
\end{equation}
with $\overline{F}(S_{\Psi^{\pm}})$ and $\overline{F}(S_{\Phi^{\pm}})$ given 
by Eqs.~(\ref{f1}) and (\ref{f4}).

Equation (\ref{fmax2}) is the first teleportation based QCP detector we will 
be using in this work. The second one is based on the trace distance. 

If we repeat the above calculation using Eqs.~(\ref{Dj2}), (\ref{qpsi}) and (\ref{qphi}), the mean trace distance given by Eq.~(\ref{Dbar}) becomes, 
\begin{eqnarray}
\overline{D}(S_{\Psi^{\pm}}) 
&=& \frac{1}{4}\left[(2 + z^2 + zz) |z| + |z^3 - z\cdot zz|\right], \label{d1} \\ 
\overline{D}(S_{\Phi^{\pm}}) 
&=& \frac{1}{4}\left[(2 - z^2 - zz) |z| + |z^3 - z\cdot zz|\right], \label{d4}
\end{eqnarray}
and the minimum mean trace distance is according to Eq.~(\ref{dmin}) 
given by
\begin{eqnarray}
\overline{\mathcal{D}} &=& \min{\left[\,\overline{D}(S_{\Psi^{\pm}}),
\overline{D}(S_{\Phi^{\pm}})\right] } \nonumber \\
&=& \frac{1}{4}\left[(2 - |z^2 + zz|) |z| + |z^3 - z\cdot zz|\right].
\label{dmin2}
\end{eqnarray}
Equation (\ref{dmin2}) is our second teleportation based QCP detector.
In Secs. \ref{secXXZ} and \ref{secXY} we investigate the effectiveness of 
both QCP detectors, Eqs.~(\ref{fmax2}) and (\ref{dmin2}), in spotlighting the 
correct location of the QCP for several models at zero and non-zero temperatures.

\textit{Remark 3:} Equations (\ref{fmax2}) and (\ref{dmin2}) do not depend on
the two-point correlation functions $xx=\med{\sigma_j^x\sigma_{j+1}^x}$ and
$yy=\med{\sigma_j^y\sigma_{j+1}^y}$, as can be seen by looking at Eqs.~(\ref{f1}),
(\ref{f4}), ({\ref{d1}}), and (\ref{d4}). The underlying reason for this is
the specific form of $\rho_1$ [Eq.~(\ref{rho1})]. Whenever the input state
to be teleported is diagonal in the standard basis
$\{|0\rangle, |1\rangle\}$, the expressions for the fidelity and for the 
trace distance will not depend on $xx$ and $yy$ if $\rho_{23}$ is 
given by Eq.~(\ref{rhoAB2}). For input states having
non-diagonal terms, though, a direct calculation shows that 
both the fidelity and the trace distance will also depend on $xx$ and $yy$.

\textit{Remark 4:} Contrary to the scenario in which the qubit to be teleported
is external to the spin chain \cite{pav23}, the expressions for the fidelity 
and for the trace distance now depend on the 
one-point correlation function $z=\med{\sigma_j^z}$. This happens because in
the present case (internal teleportation), the input state to be teleported 
is a function of $z$, as can be seen by looking at Eq.~(\ref{rho1}). And this 
implies that the
teleported states with Bob are, according to Eq.~(\ref{stepD}),
\begin{eqnarray}
\rho_{_{B_{\Psi^{\pm}}}}(S_{\Psi^{\pm}})&=& \frac{1}{2(1-z^2)} 
\left(\hspace{-.15cm}
\begin{array}{cc}
g(zz,-z) & 0\\
0 &  g(zz,z)
\end{array}
\hspace{-.15cm}\right), \label{bob1}
\\
\rho_{_{B_{\Phi^{\pm}}}}(S_{\Psi^{\pm}})&=& \frac{1}{2(1+z^2)} 
\left(\hspace{-.15cm}
\begin{array}{cc}
f(zz,-z) & 0\\
0 &  f(zz,z)
\end{array}
\hspace{-.15cm}\right), \label{bob2}
\\
\rho_{_{B_{\Psi^{\pm}}}}(S_{\Phi^{\pm}})&=& \frac{1}{2(1-z^2)} 
\left(\hspace{-.15cm}
\begin{array}{cc}
g(zz,z) & 0\\
0 &  g(zz,-z)
\end{array}
\hspace{-.15cm}\right), \label{bob3}
\\
\rho_{_{B_{\Phi^{\pm}}}}(S_{\Phi^{\pm}})&=& \frac{1}{2(1+z^2)} 
\left(\hspace{-.15cm}
\begin{array}{cc}
f(zz,z) & 0\\
0 &  f(zz,-z)
\end{array}
\hspace{-.15cm}\right), \label{bob4}
\end{eqnarray}
where $f(zz,z)$ and $g(zz,z)$ are given by Eqs.~(\ref{f}) and (\ref{g}), 
respectively.
In Eqs.~(\ref{bob1})-(\ref{bob4}) we have explicitly written the dependence of
Bob's final state on the set $S_k$  
containing the unitary corrections that he can implement on his
qubit. In other words, $\rho_{_{B_j}}(S_k)$ denotes the final state with Bob if
Alice measures the Bell state $j$ and Bob corrects his qubit using the 
corresponding unitary correction listed in $S_k$ [cf. Eqs.~(\ref{s1})-(\ref{s4})].

\textit{Remark 5:} If we set $z=0$, Eqs.~(\ref{f1}),
(\ref{f4}), ({\ref{d1}}), and (\ref{d4}) become 
$\overline{F}(S_{\Psi^{\pm}})=\overline{F}(S_{\Phi^{\pm}})=1$ and
$\overline{D}(S_{\Psi^{\pm}})=\overline{D}(S_{\Phi^{\pm}})=0$. This means that
the present method to detect QCPs does not work for the classes of 
spin chains studied here that do not have a 
net magnetization. For instance, it does not work for the XXZ model when the 
external magnetic field is zero since in this case we always have $z=0$ 
as we drive the system across its QCPs \cite{wer10b,pav23}. In other words,
whenever $z=0$ the output state with Bob at the end of the 
teleportation protocol will always be 
exactly equal to the input state of Alice regardless of the quantum phase 
of the system. We will always obtain the same constant
values for the maximum mean fidelity ($\overline{\mathcal{F}}=1$) and for the
minimum mean trace distance ($\overline{\mathcal{D}}=0$) in all quantum phases.

\section{The XXZ model in an external field}
\label{secXXZ}

The XXZ spin chain in an external longitudinal magnetic field is described by the 
following Hamiltonian, where we set $\hbar=1$,
\begin{equation}
H = \sum_{j=1}^{L}\left(\sigma^{x}_{j}\sigma^{x}_{j+1} +
\sigma^{y}_{j}\sigma^{y}_{j+1} + \Delta
\sigma^{z}_{j}\sigma^{z}_{j+1} - \frac{h}{2}\sigma^z_j\right). \label{Hxxz}
\end{equation}
In Eq.~(\ref{Hxxz}) $\Delta$ is the tuning parameter and $h$ is the external 
field.

The XXZ model has two QCPs at zero temperature \cite{yan66,clo66,klu92,bor05,boo08,tri10,tak99}. 
For a given field $h$, as we vary $\Delta$ we see at  
$\Delta=\Delta_1$ the first QPT, with the system's ground
state changing from a ferromagnetic ($\Delta < \Delta_1$) to a critical 
antiferromagnetic phase ($\Delta_1 < \Delta < \Delta_2$). 
If we continue to increase $\Delta$, we have at $\Delta=\Delta_2$ 
another QPT, where the spin chain becomes an Ising-like antiferromagnet.

The value of $\Delta_1$ as a function of $h$ is given by the solution of 
\begin{equation}
h=4J(1+\Delta_1).
\label{delta1}
\end{equation}
On the other hand, $\Delta_2$ is obtained by solving the following equation,
\begin{equation}
h=4\sinh(\eta)\sum_{j=-\infty}^\infty\frac{(-1)^j}{\cosh(j\eta)},
\label{dinf}
\end{equation}
with $\eta = \cosh^{-1}(\Delta_2)$.
For $h=12$, the value of $h$ adopted in this work, 
the two critical points are
\begin{eqnarray}
\Delta_1 &=& 2.000, \\ 
\Delta_2 &=& 4.875,
\end{eqnarray}
with the latter QCP correct within a numerical error of $\pm 0.001$.

\begin{figure}[!ht]
\includegraphics[width=8cm]
{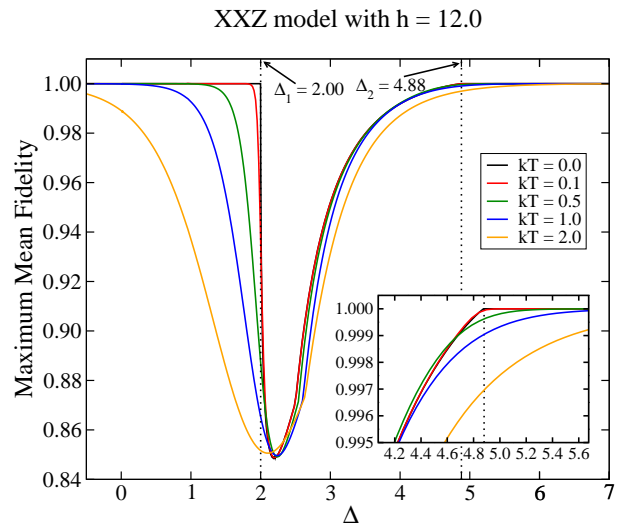}
\caption{\label{fig_max_geral}(color online) $\overline{\mathcal{F}}$, Eq.~(\ref{fmax2}), as a function of
$\Delta$ for $h=12.0$. 
At $T=0$ both QCPs are detected by a discontinuity in
the derivative of $\overline{\mathcal{F}}$ with respect to $\Delta$.
See the inset for a better appreciation of this fact at the second QCP.
For $T>0$, these discontinuities in the derivatives are smoothed out and
manifest themselves in high values for the magnitudes of the derivatives
around the critical points. The dotted lines mark the 
QCPs and for the solid curves the temperature increases from top to bottom
when $\Delta<\Delta_1$. 
Here and in all other graphs all quantities are dimensionless.}
\end{figure}

Let us start studying the efficacy of the maximum mean fidelity
in detecting the two QCPs for the XXZ model in an external field. 
In Fig. \ref{fig_max_geral} we show $\overline{\mathcal{F}}$ 
as a function of $\Delta$ for $h=12$ and several values of $T$.

Looking at Fig. \ref{fig_max_geral}, we realize that $\overline{\mathcal{F}}$ 
clearly detects the first QCP, with an efficacy compatible to the one 
we have when employing the external teleportation approach of Ref. \cite{pav23}.
The detection of the second QCP, though, is not good enough. As we see in the 
inset of Fig. \ref{fig_max_geral}, we need to know the fidelity within an accuracy
of $0.001$ to observe a discontinuity in its derivative. And as we increase $T$,
we rapidly lose any means of tracking the correct location of the second QCP.

\begin{figure}[!ht]
\includegraphics[width=8cm]{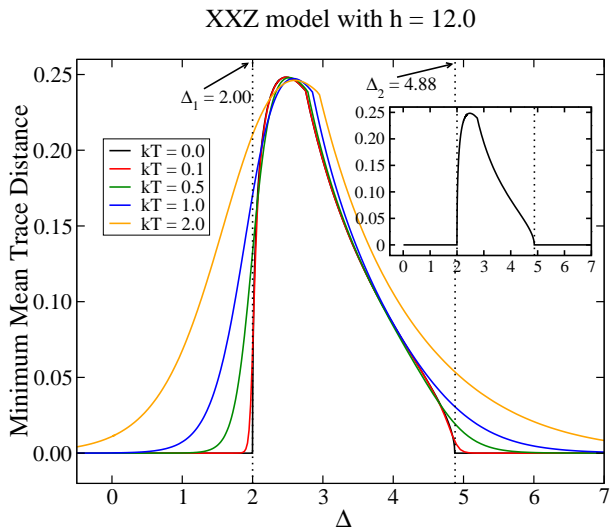}
\caption{\label{fig_dmin_geral}(color online) $\overline{\mathcal{D}}$, Eq.~(\ref{dmin2}), as a function of
$\Delta$ for $h=12.0$. 
Note that now at $T=0$ (see inset)
both QCPs are clearly detected by a discontinuity in
the derivative of $\overline{\mathcal{D}}$ with respect to $\Delta$.
For $T>0$, there are no longer discontinuities in the derivatives. In this case
we have
high values for the derivatives about the location of the discontinuities that
occurred at $T=0$. For $kT \lesssim 0.5$, the
extrema of the derivatives lie close together and by extrapolating to 
$kT \rightarrow 0$ we can correctly determine the QCPs. The dotted lines mark the
exact locations of the QCPs and for the solid curves the temperature increases from bottom to top when $\Delta<\Delta_1$ and $\Delta>\Delta_2$.}
\end{figure}

This problem is solved by working with the trace distance. In Fig. \ref{fig_dmin_geral} we show $\overline{\mathcal{D}}$ 
as a function of $\Delta$ for $h=12$ and the same values of $kT$ given in 
Fig. \ref{fig_max_geral}. 

Looking at Fig. \ref{fig_dmin_geral} we note that the two QCPs ($\Delta_1$ 
and $\Delta_2$) are clearly detected by discontinuities in the derivatives of 
$\overline{\mathcal{D}}$ with
respect to $\Delta$ as we go through the critical points. 
Also, in contrast to the behavior of $\overline{\mathcal{F}}$ in the external teleportation approach \cite{pav23}, we do not see two extra discontinuities in the derivatives of
$\overline{\mathcal{F}}$ and of $\overline{\mathcal{D}}$ 
[cf. Figs. \ref{fig_max_geral} and \ref{fig_dmin_geral} with the 
corresponding ones given in Ref. \cite{pav23}]. In the present case,
we only have one small kink between the two QCPs.

\begin{figure}[!ht]
\includegraphics[width=8cm]{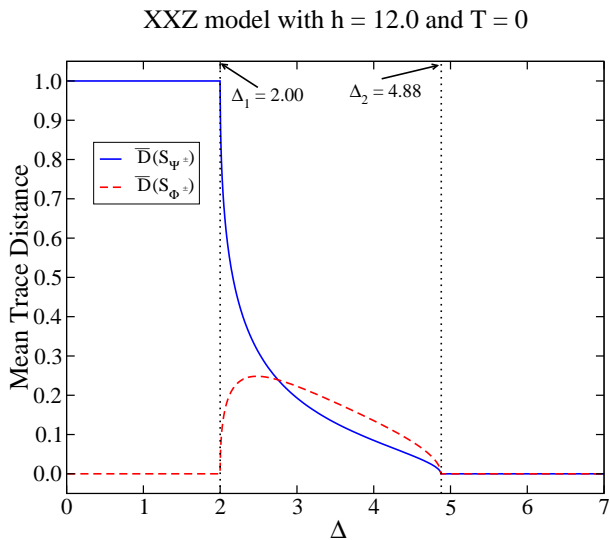}
\caption{\label{fig_dmin1e2}(color online) $\overline{D}(S_k)$,
Eqs.~(\ref{d1}) and (\ref{d4}), as a function of $\Delta$ when
$T=0$ and $h=12.0$.}
\end{figure}

This extra kink is related to the exchange of the functions maximizing 
$\overline{\mathcal{F}}$ or minimizing $\overline{\mathcal{D}}$ as we
reach the location of that extra kink. In Fig. \ref{fig_dmin1e2} we illustrate this point showing at $T=0$ 
the two expressions appearing in the definition of  
$\overline{\mathcal{D}}$, Eqs.~(\ref{d1}) and (\ref{d4}), as functions of 
$\Delta$.
Before the extra kink the expression 
minimizing $\overline{\mathcal{D}}$ is
given by $\overline{D}(S_{\Phi^{\pm}})$ and after it we have 
$\overline{D}(S_{\Psi^{\pm}})$. The point where $\overline{D}(S_{\Phi^{\pm}})$
and $\overline{D}(S_{\Psi^{\pm}})$ intersect each other 
is exactly the point where we see
the extra kink in the curve of $\overline{\mathcal{D}}$. 
Note also that the location of this extra kink changes as we increase the temperature and that the same analysis applies to the fidelity and to higher temperatures as well.

To estimate the values of the two QCPs using finite $T$ data, 
we employ the same techniques of Refs. \cite{wer10b,pav23}. 
For finite $T$, the $T=0$ discontinuities in the 
derivatives of $\overline{\mathcal{D}}$ with respect to $\Delta$ at the QCPs 
manifest themselves in high values for the magnitudes of these derivatives
around the QCPs. 
The location of these 
extremum values around the QCPs for several values of $T$
are the key finite $T$ data we need to obtain the QCPs. 
We can predict the values of the two QCPs by extrapolating to $T=0$  
how the locations of those extremum values change 
as we decrease the temperature.

\begin{figure}[!ht]
\includegraphics[width=8cm]{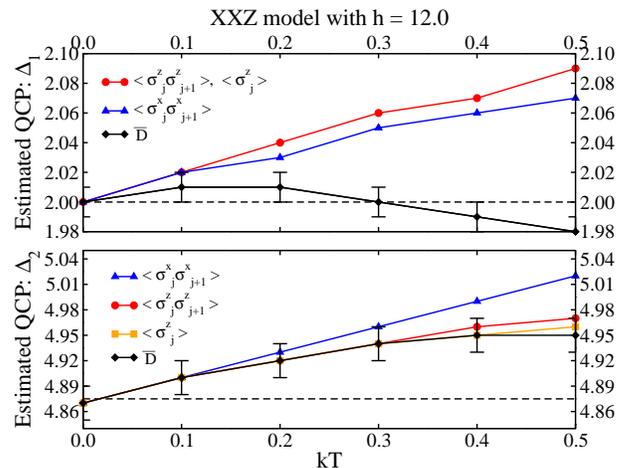}
\caption{\label{fig_qcpT} (color online) Estimated QCPs  
after determining the spots of the 
extrema of the first order (upper panel) and 
of the second order (lower panel) derivatives with respect to $\Delta$ 
for $\overline{\mathcal{D}}, 
\langle \sigma^x_j\sigma^x_{j+1} \rangle, \langle \sigma^z_j\sigma^z_{j+1} \rangle$, and $\langle \sigma^z_j\rangle$ at several different values of $T$. 
The dashed lines mark the exact values of the QCPs. 
The error bars in the curves showing the locations of the 
extrema of the derivatives of  
$\overline{\mathcal{D}}$ represent
the numerical uncertainty in determining their exact location.
For the other quantities in the two panels, similar error bars (not shown) exist
about the dots showing the locations of the extrema of their derivatives.
See text for details.}
\end{figure}

Specifically, for six different temperatures, namely, 
$kT=0, 0.1, 0.2, 0.3, 0.4$, and $0.5$,
we calculated around the two critical points the values of
the one- and two-point correlation functions and of $\overline{\mathcal{D}}$ as a function of $\Delta$. We changed $\Delta$ in increments
of $0.01$. Afterwards, we numerically evaluated the first order derivatives of 
those quantities about $\Delta_1$ and their second order derivatives about 
$\Delta_2$. The 
highest values for the magnitudes of those derivatives can be seen in 
Fig. \ref{fig_qcpT}. Moreover, taking into account that $\Delta$ was changed in 
increments of $0.01$, the location of the 
maxima of the magnitudes of the first order derivatives are obtained 
with an error of $\pm 0.01$. On the other hand, the second order derivatives 
are calculated from the first order ones, which already possess an error of 
$\pm 0.01$. 
Therefore, we estimate that the error of the correct spot for the extremum 
values of the second order derivatives are no less than $\pm 0.02$. 
To not clutter the visualization of the other 
curves, in Fig. \ref{fig_qcpT} we draw error bars depicting these uncertainties 
only for $\overline{\mathcal{D}}$. 

Looking at Fig. \ref{fig_qcpT}, we note that for all temperatures
the extremum of the first order derivative of $\overline{\mathcal{D}}$ gives the best approximation for the exact value of 
the first QCP (upper panel of Fig. \ref{fig_qcpT}). 
The second QCP is also better approximated by the the extremum of the 
second order derivative of $\overline{\mathcal{D}}$ 
(lower panel of Fig. \ref{fig_qcpT}). 
However, as we decrease $kT$ the other quantities shown in Fig. \ref{fig_qcpT} 
catch up. Around $kT=0.1$ they are all equally efficient in estimating the location of the second QCP. Also, remembering that for the external teleportation strategy
the behavior of the derivatives of $\overline{\mathcal{F}}$ is equal to the 
behavior of the derivatives of 
$\langle \sigma^z_j\sigma^z_{j+1}\rangle$ \cite{pav23}, we have that 
the internal teleportation strategy gives better 
estimates for the location of the two QCPs when compared with the external 
teleportation approach.
 
Excluding the $kT=0$ point, 
we can implement linear and quadratic regressions to find the best curves 
fitting the remaining data ($kT=0.1, 0.2, 0.3, 0.4, 0.5$) \cite{pav23}. 
For all quantities given in the lower panel of Fig. \ref{fig_qcpT},
a linear regression is enough to predict 
the correct location of the QCP $\Delta_2$ as we take the $T=0$ limit.
The predicted location of $\Delta_2$ is given within an accuracy of $\pm 0.02$,
the estimated numerical error for the locations of the extrema of the second 
order derivatives of all the quantities given in Fig. \ref{fig_qcpT}. 
The same conclusion 
applies for all but one of the curves given in the upper panel of 
Fig. \ref{fig_qcpT}, where again the fitted straight lines give the correct 
location of the QCP $\Delta_1$ within an accuracy of $\pm 0.01$, 
the numerical error for the locations of the extrema of the first order derivatives of all the quantities given in Fig. \ref{fig_qcpT}. To achieve
this same level of accuracy for $\overline{\mathcal{D}}$, though, we need a 
quadratic regression.

\textit{Remark 6:} Although the results reported in this section were obtained
assuming that the external magnetic field was $h=12$, similar results are 
obtained for other values of the field.

\textit{Remark 7:} Instead of using $\overline{\mathcal{D}}$ [Eq.~(\ref{dmin2})]
and its derivatives to estimate the QCPs using finite $T$ data,
we could have worked directly with either $\overline{D}(S_{\Phi^{\pm}})$
or $\overline{D}(S_{\Psi^{\pm}})$ [Eq.~(\ref{d1}) or (\ref{d4})] and their
derivatives. All three quantities give similar results in predicting the location of the QCPs.

\section{The XY and the Ising model}
\label{secXY}

A spin-1/2 chain in an external transverse magnetic field described by 
the XY model is given by the following Hamiltonian \cite{lie61,bar70,bar71},
\begin{equation}
H \!=\! -\frac{\lambda}{4}\!\sum_{j=1}^{L}\!\left[(1+\gamma)\sigma^{x}_{j}\sigma^{x}_{j+1} + (1-\gamma)\sigma^{y}_{j}\sigma^{y}_{j+1}\right]\! 
- \frac{1}{2}\!\sum_{j=1}^{L}\!\sigma^z_j. \label{Hxy}
\end{equation}
In Eq.~(\ref{Hxy}) $\lambda$ is associated with the inverse of the strength of 
the magnetic field and $\gamma$ is the anisotropy parameter. 
We obtain the Ising transverse model from Eq.~(\ref{Hxy}) by setting 
$\gamma=\pm 1$. And if $\gamma=0$ we have the XX model in a transverse magnetic
field.

The tuning parameter for the present model can be either $\lambda$ or $\gamma$.
Fixing $\gamma$, we have that 
for $\lambda < 1$ the system is in a ferromagnetic ordered
phase and as we increase $\lambda$ we reach the QCP $\lambda_c=1.0$,
where the Ising transition occurs. For $\lambda > 1$ the system now lies
in a quantum paramagnetic phase \cite{pfe70}.
There is another QPT for this model when $\lambda > 1$ if we change
the anisotropy parameter $\gamma$ instead of $\lambda$. 
This ``anisotropy transition'' 
occurs at $\gamma_c=0$ \cite{lie61,bar70,bar71,zho10}, where 
the critical point $\gamma_c$ separates a ferromagnet ordered phase 
in the $x$-direction from a ferromagnet ordered phase in the 
$y$-direction.

\subsection{The $\lambda$ transition}

\begin{figure}[!ht]
\includegraphics[width=8cm]{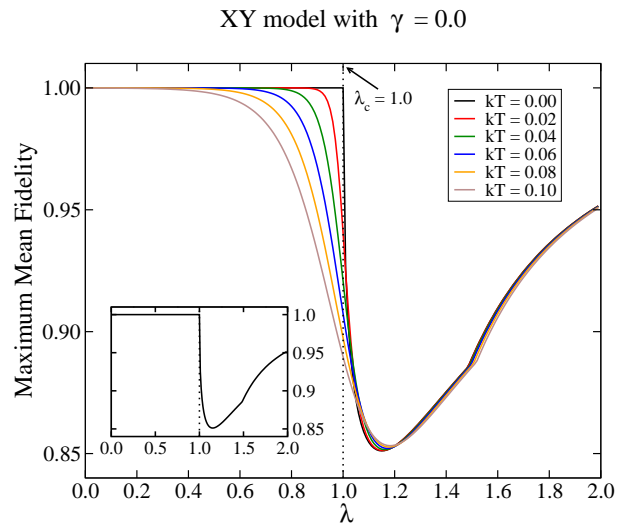}
\caption{\label{fig_max_geral1}(color online) $\overline{\mathcal{F}}$, Eq.~(\ref{fmax2}), as a function of
$\lambda$ and for several values of $kT$ when we fix 
$\gamma=0.0$ (XX model in a transverse field) 
[see Eq.~(\ref{Hxy})]. The inset highlights the $T=0$ curve.
The dotted lines mark the 
QCP $\lambda_c$ and for the solid curves the temperature increases from top to bottom when $\lambda < \lambda_c$.}
\end{figure}

\begin{figure}[!ht]
\includegraphics[width=8cm]{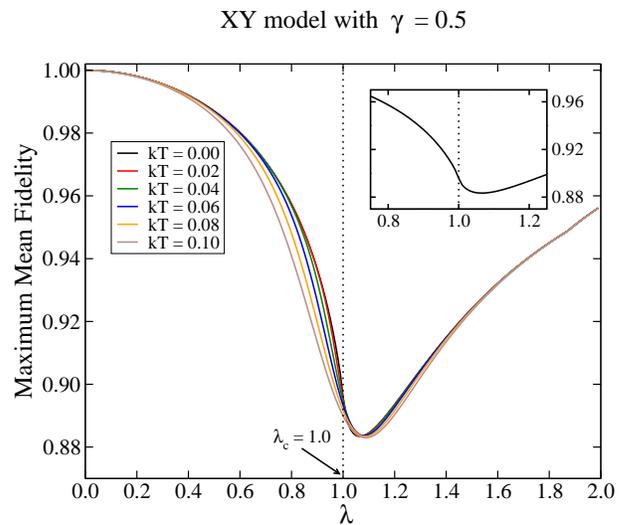}
\caption{\label{fig_max_geral2}(color online) Same as Fig. \ref{fig_max_geral1}
but now $\gamma=0.5$. The dotted lines mark the 
QCP $\lambda_c$ and for the solid curves the temperature increases from top to bottom when $\lambda < \lambda_c$.}
\end{figure}

\begin{figure}[!ht]
\includegraphics[width=8cm]{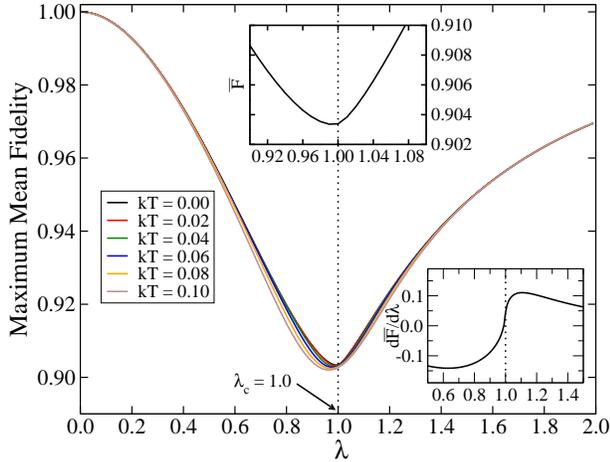}
\caption{\label{fig_max_geral3}(color online) Same as Fig. \ref{fig_max_geral1}
but now $\gamma=1.0$ (Ising model in a transverse field).
The upper inset highlights the $T=0$ curve of $\overline{\mathcal{F}}$ and
the lower inset its derivative with respect to $\lambda$.
The dotted lines mark the 
QCP $\lambda_c$ and for the solid curves the temperature increases from top to bottom when $\lambda < \lambda_c$.}
\end{figure}

In Figs. \ref{fig_max_geral1}, \ref{fig_max_geral2}, and 
\ref{fig_max_geral3} we plot the maximum mean fidelity 
$\overline{\mathcal{F}}$ as a function of
$\lambda$ for three values of $\gamma$ and for several values of temperature.
The chosen values for $\gamma$ are such that we obtain the isotropic
XX model ($\gamma=0.0$), the anisotropic XY model ($\gamma=0.5$), and
the Ising model ($\gamma=1.0$).

For the isotropic XX model ($\gamma=0.0$), the QCP is clearly detect 
at $T=0$ by a discontinuity in the derivative of $\overline{\mathcal{F}}$ at
$\lambda_c=1.0$ (see Fig. \ref{fig_max_geral1}) as well as discontinuities 
in the derivatives of the one- and two-point correlation functions \cite{pav23}. 
For $T>0$, the discontinuity in the derivative manifests itself
in a high value for the derivative's magnitude about the QCP.  
The location of the extremum values of the derivatives move away from the 
QCP as we change $T$. Nevertheless, as we will see for $kT \lesssim 0.1$, these
extrema lie close to the correct value of the QCP and we can properly infer
the correct location of $\lambda_c$ by extrapolating to $kT=0$.

When $\gamma=0.5$ (anisotropic XY model) and $\gamma=1.0$ 
(Ising transverse model), the
QCP is determined, respectively, by the inflection points of 
$\overline{\mathcal{F}}$ and $d\overline{\mathcal{F}}/d\lambda$ 
that occur exactly at $\lambda_c=1.0$
when $T=0$ (see Figs. \ref{fig_max_geral2} and \ref{fig_max_geral3}).
For higher values of $T$, the inflection points are displaced away 
from $\lambda_c$ and the best strategy to estimate the QCP is  
by determining the location of the maximum (minimum) of the second order 
derivatives of $\overline{\mathcal{F}}$ with respect to $\lambda$.
The location of these extremum values lie close to $\lambda_c$ when $kT \lesssim 0.1$ and, as we will show below, by extra\-po\-lating to the absolute zero temperature we can predict its correct value.
Note also that the one- and two-point correlation functions have inflection 
points at the QCP when $T=0$ \cite{pav23}.

\begin{figure}[!ht]
\includegraphics[width=8cm]{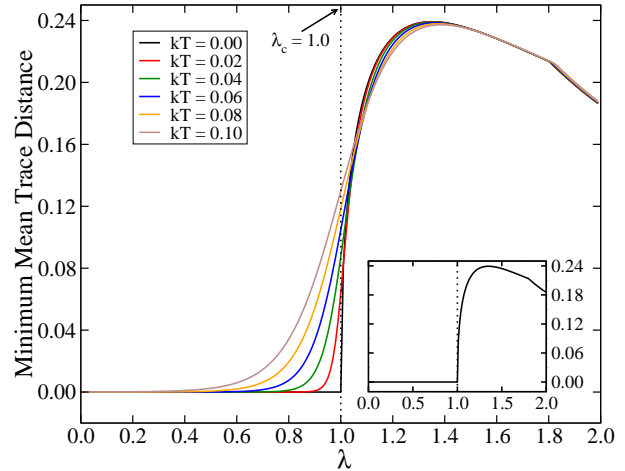}
\caption{\label{fig_min_geral1}(color online) $\overline{\mathcal{D}}$, Eq.~(\ref{dmin2}), as a function of
$\lambda$ for several values of $kT$ while fixing 
$\gamma=0.0$ (XX model in a transverse field). 
The inset shows the $T=0$ curve and the dotted lines mark the 
QCP $\lambda_c$. For the solid curves the temperature decreases from top to bottom when $\lambda < \lambda_c$.}
\end{figure}

\begin{figure}[!ht]
\includegraphics[width=8cm]{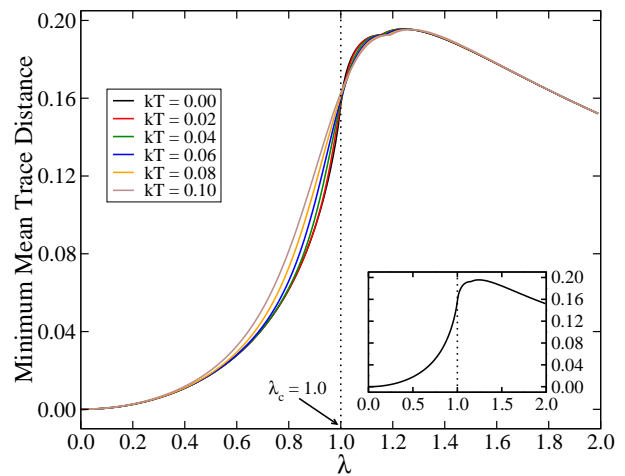}
\caption{\label{fig_min_geral2}(color online) Same as Fig. \ref{fig_min_geral1}
but now $\gamma=0.5$. The inset shows the $T=0$ curve and the dotted lines mark the 
QCP $\lambda_c$. For the solid curves the temperature decreases from top to bottom when $\lambda < \lambda_c$.}
\end{figure}

\begin{figure}[!ht]
\includegraphics[width=8cm]{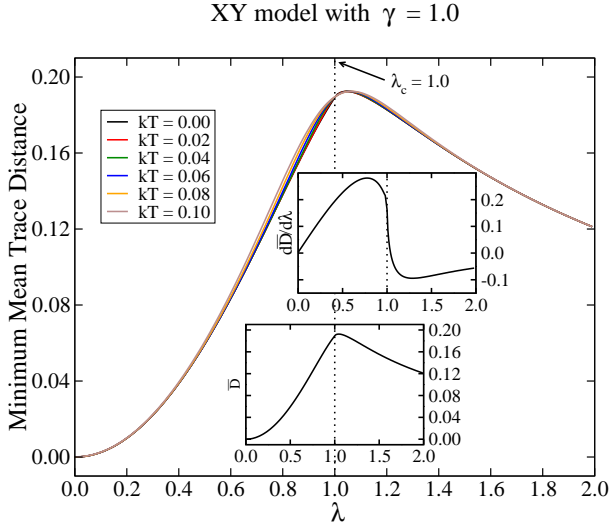}
\caption{\label{fig_min_geral3}(color online) Same as Fig. \ref{fig_min_geral1}
but now $\gamma=1.0$ (Ising model in a transverse field).
The lower inset highlights the $T=0$ curve of $\overline{\mathcal{D}}$ and
the upper inset its derivative with respect to $\lambda$.
The dotted lines mark the 
QCP $\lambda_c$. For the solid curves the temperature decreases from top to bottom when $\lambda < \lambda_c$.}
\end{figure}

In Figs. \ref{fig_min_geral1}, \ref{fig_min_geral2}, and 
\ref{fig_min_geral3} we plot the minimum mean trace distance 
$\overline{\mathcal{D}}$ as a function of
$\lambda$ for the same three values of $\gamma$ and temperatures shown 
in Figs. \ref{fig_max_geral1} to \ref{fig_max_geral3}. 
Looking at Figs. \ref{fig_max_geral1} to \ref{fig_min_geral3}, and taking into 
account that a high value for $\overline{\mathcal{F}}$ implies a low value for
$\overline{\mathcal{D}}$,  we realize that
the behavior of $\overline{\mathcal{F}}$ and $\overline{\mathcal{D}}$ about the
QCPs are very similar for a given value of $\gamma$.

Similarly to the behavior of $\overline{\mathcal{F}}$ at $T=0$, for 
the isotropic XX model ($\gamma=0.0$) the QCP is identified by 
a discontinuity in the derivative of $\overline{\mathcal{D}}$
with respect to $\lambda$ at $\lambda_c$ (see Fig. \ref{fig_min_geral1}). 
For $T>0$, we no longer have a discontinuity in the derivative of 
$\overline{\mathcal{D}}$ but rather a
high value for its magnitude about the QCP.  
The location of this extremum moves away from the exact value of the 
QCP as we increase $T$.  For $kT \lesssim 0.1$, though, these
extrema are near to the correct value of the QCP and 
we will show in what follows how
to use these finite $T$ data to extract the 
exact location of the QCP.

The same qualitative features observed for the behavior of
$\overline{\mathcal{F}}$ as a function of $\lambda$ can be seen 
for $\overline{\mathcal{D}}$ when we work with $\gamma=0.5$ 
(anisotropic XY model) and 
$\gamma=1.0$ (Ising transverse model). Again, at $T=0$ the 
critical point $\lambda_c$
is given, respectively, by an inflection point of 
$\overline{\mathcal{D}}$ and of $d\overline{\mathcal{D}}/d\lambda$ 
at $\lambda_c=1.0$ (see Figs. \ref{fig_min_geral2} and \ref{fig_min_geral3}). 
If we increase $T$, though, the inflection points move away 
from $\lambda_c=1.0$. As we will see, using finite $T$ data 
we can best estimate the correct value for 
$\lambda_c$ by computing the location of the extremum values of the second order 
derivatives of $\overline{\mathcal{D}}$ with respect to $\lambda$ and then 
numerically taking the $T=0$ limit.

\begin{figure}[!ht]
\includegraphics[width=8cm]{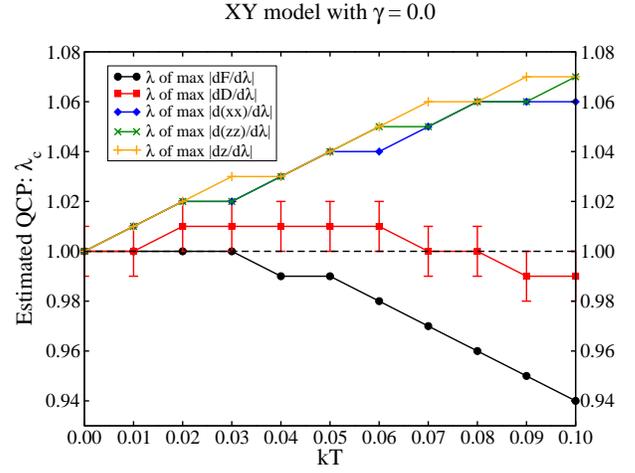}
\caption{\label{fig_qcpTxy1}(color online) Estimated value for the critical point  
$\lambda_c$ using finite $T$ data. The locations of the maxima of the first order derivatives of $\overline{\mathcal{D}}$ with respect to $\lambda$
for several values of $T$ best approximate the correct location of the 
QCP (red-square curve). The error bars depict the uncertainties in
numerically obtaining the location of the maxima of 
$d\overline{\mathcal{D}}/d\lambda$. 
The dashed line gives the exact value of the QCP and
the other solid 
curves give the locations of the extrema of the first order derivatives of
$\overline{\mathcal{F}}$ and of the one- and two-point correlation functions.
For ease of visualization, we do not show the error bars in these cases.
See text for details.}
\end{figure}

Let us show now how to determine the QCPs using finite $T$
data. Following the strategy outlined in Sec. \ref{secXXZ}, for several 
values of temperature we numerically 
evaluate the first and second order derivatives  with respect to $\lambda$ of 
$\overline{\mathcal{D}}$, $\overline{\mathcal{F}}$, and of the one- and two-point 
correlation functions. Then, we employ
the location of the extrema of those derivatives as the value for the QCP. 
When $\gamma=0.0$, the location of the maximum of
the first order derivative of $\overline{\mathcal{D}}$ is the optimal strategy
to obtain the correct value for $\lambda_c$ (see Fig. \ref{fig_qcpTxy1}). Indeed,
for 
all values of temperature up to $kT=0.1$, the predicted value for $\lambda_c$
lies within the correct $T=0$ value ($\lambda_c=1.0$) if we take into account,
as explained in Sec. \ref{secXXZ},
the uncertainties ($\pm 0.01$) in obtaining the spot of the maximum of 
$d\overline{\mathcal{D}}/d\lambda$.  

\begin{figure}[!ht]
\includegraphics[width=8cm]{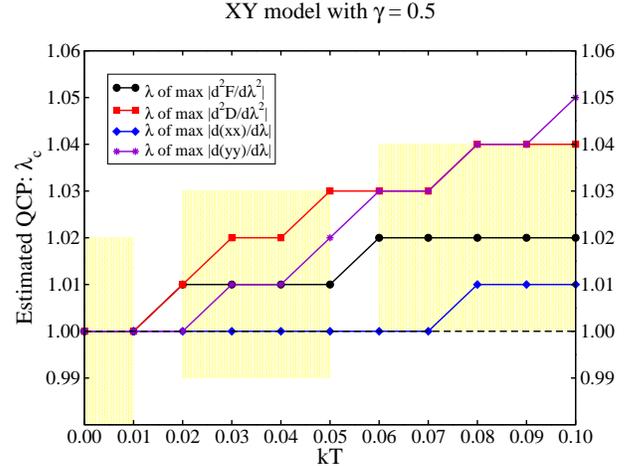}
\caption{\label{fig_qcpTxy2}(color online) Same as Fig. \ref{fig_qcpTxy1}. 
The yellow-shaded regions represent the uncertainty in the 
numerical determination of the locations of the extrema for the 
second derivatives of $\overline{\mathcal{F}}$ ($\pm 0.02$).
For ease of visualization, we do not show the errors associated with the other
quantities and we do not show the curves and points 
giving the locations of the extrema of $dz/d\lambda$
and $d(zz)/d\lambda$. The latter curves lie between the curves above 
depicting the locations of the extrema for $d(yy)/d\lambda$ and $d(xx)/d\lambda$.}
\end{figure}

If we now fix $\gamma=0.5$, the extrema of $d^2\overline{\mathcal{F}}/d\lambda^2$
lie closer to the true value of the QCP than those of 
$d^2\overline{\mathcal{D}}/d\lambda^2$ when $kT>0.02$ 
(see Fig. \ref{fig_qcpTxy2}). 
For high values of $kT$, the extrema
of $d(xx)/d\lambda$ are the optimal choice to estimate the QCP and, as we keep
decreasing $T$, all quantities eventually give the exact location for the QCP.
It is worth noting that if we take into account the errors (yellow-shaded regions
in Fig. \ref{fig_qcpTxy2}) in numerically locating
the extrema of $d^2\overline{\mathcal{F}}/d\lambda^2$, for all temperatures
shown in Fig. \ref{fig_qcpTxy2} the exact value
of the QCP lies within the range of possible values for the location of 
those extrema.

\begin{figure}[!ht]
\includegraphics[width=8cm]{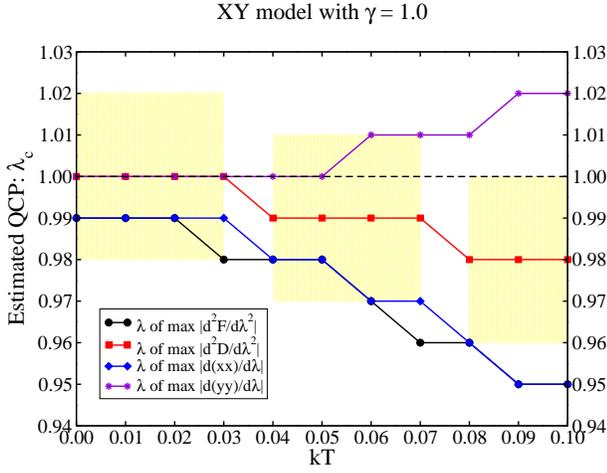}
\caption{\label{fig_qcpTxy3}(color online) Same as Fig. \ref{fig_qcpTxy1}. 
The yellow-shaded regions represent the uncertainty in the 
numerical determination of the locations of the extrema for the 
second derivatives of $\overline{\mathcal{D}}$ ($\pm 0.02$).
For ease of visualization, we do not show the errors associated with the other
quantities and we do not show the locations of the extrema of $dz/d\lambda$
and $d(zz)/d\lambda$, which lie between the curves above showing the 
locations of the extrema for $d(yy)/d\lambda$ and $d(xx)/d\lambda$.}
\end{figure}

When $\gamma=1.0$, we realize looking at Fig. \ref{fig_qcpTxy3} that 
up to $kT=0.03$ the extrema of $d(yy)/d\lambda$ and 
$d^2\overline{\mathcal{D}}/d\lambda^2$ give the correct value for the QCP.
And if we take into account the errors in numerically estimating the 
extrema of $d^2\overline{\mathcal{D}}/d\lambda^2$, both quantities are 
almost equally optimal in predicting the correct value of the QCP 
for higher values of $kT$ (see Fig. \ref{fig_qcpTxy3}). 

We can be more quantitative in the prediction of the location of the QCP by 
following the same strategy of Ref. \cite{pav23} that was already explained 
in Sec. \ref{secXXZ}. For all the quantities shown in Figs. \ref{fig_qcpTxy1} to
\ref{fig_qcpTxy3}, we excluded the $kT=0.0$ point and implemented 
a linear regression with the remaining data 
($kT=0.01, 0.02, \ldots, 0.10$). Using the fitted curves, 
we took the $kT=0$ limit and obtained for all cases and 
within the numerical errors already reported the correct value for the QCP.

\subsubsection*{A closer look at the curves for $\overline{\mathcal{D}}$}

\begin{figure}[!ht]
\includegraphics[width=8cm]{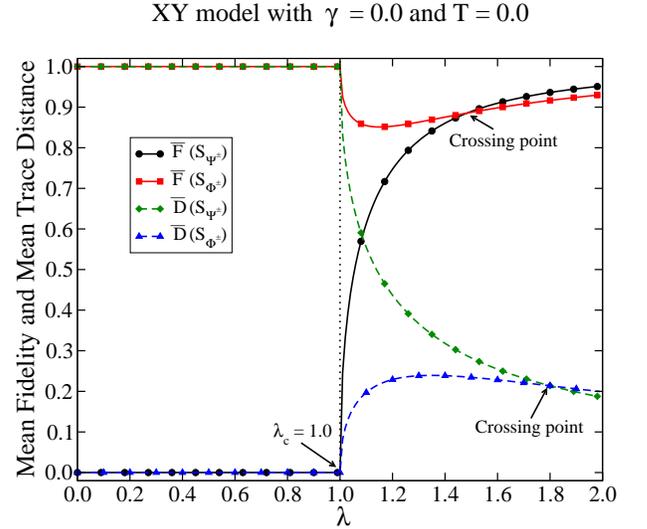}
\caption{\label{fig_extra1}(color online) Mean fidelities 
$\overline{F}(S_{\Psi^{\pm}})$ and $\overline{F}(S_{\Phi^{\pm}})$ 
[Eqs. (\ref{f1}) and (\ref{f4})] and mean trace distances 
$\overline{D}(S_{\Psi^{\pm}})$ and $\overline{D}(S_{\Phi^{\pm}})$ 
[Eqs. (\ref{d1}) and (\ref{d4})] as a function of $\lambda$ when $kT=0$ and
$\gamma=0.0$ (XX model in a transverse field). The highlighted points,
which we called ``crossing points'', are located exactly at the cusps of 
$\overline{\mathcal{F}}$ and $\overline{\mathcal{D}}$ seen in Figs. \ref{fig_max_geral1} and \ref{fig_min_geral1}, respectively. Note that 
these cusps are not related to QPTs and they move to higher values of $\lambda$
as we increase the temperature.}
\end{figure}

Before we move to the study of the anisotropy transition, it is important to 
better analyze the behavior of the minimum mean trace distance 
$\overline{\mathcal{D}}$ as a function of $\lambda$ 
(a similar analysis applies to $\overline{\mathcal{F}}$ as well). Looking at
Fig. \ref{fig_min_geral1}, where we fixed $\gamma=0.0$, we note a cusp located
about $\lambda = 1.8$ that is not associated with a QPT. The origin of this 
cusp can be traced back to the point where the function minimizing  
$\overline{\mathcal{D}}$, namely, $\overline{D}(S_{\Phi^{\pm}})$, is replaced by 
$\overline{D}(S_{\Psi^{\pm}})$. This is illustrated in Fig. \ref{fig_extra1},
where we can also understand the cusp seen at $\lambda\approx 1.5$ for 
$\overline{\mathcal{F}}$ in Fig. \ref{fig_max_geral1}. This is the point where
$\overline{F}(S_{\Psi^{\pm}})=\overline{F}(S_{\Phi^{\pm}})$. Before this point
$\overline{\mathcal{F}}$ is given by $\overline{F}(S_{\Phi^{\pm}})$ and after
it by $\overline{F}(S_{\Psi^{\pm}})$.

\begin{figure}[!ht]
\includegraphics[width=8cm]{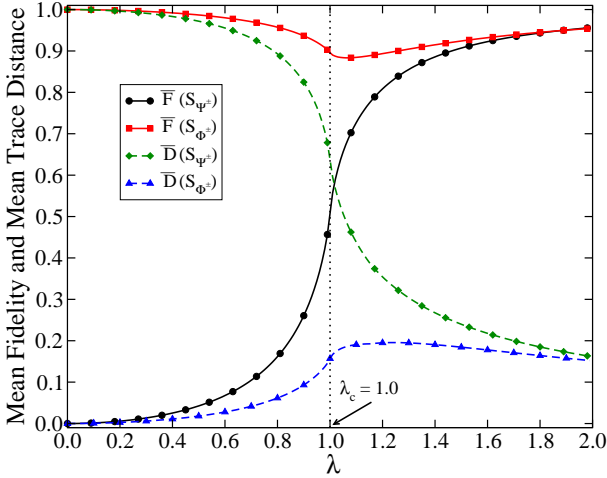}
\caption{\label{fig_extra2}(color online) Same as Fig. \ref{fig_extra1} but
now we have $\gamma=0.5$ (anisotropic XY model in a transverse field). Note
that the solid curves do not cross each other. The same applies to the dashed
ones.}
\end{figure}

Plotting the same curves for $\gamma=0.5$ and $\gamma=1.0$, we realize that 
the mean fidelity curves as well as the mean trace distance ones do not cross
each other. This is why we do not have a cusp in Figs. \ref{fig_max_geral2} and 
\ref{fig_max_geral3} and also in Fig. \ref{fig_min_geral3}. We do have a small
cusp, though, for the curves of $\overline{\mathcal{D}}$ given in Fig. \ref{fig_min_geral2}, where we fixed $\gamma=0.5$.  
This cusp is not related to a QPT and we can 
understand its origin by carefully looking at the functional form of
$\overline{D}(S_{\Phi^{\pm}})$, which gives the minimum mean trace distance 
$\overline{\mathcal{D}}$ when $0\leq \lambda \leq 2$ (see Fig. \ref{fig_extra2}).

Looking at Eq.~(\ref{d4}), we note that the second
term defining $\overline{D}(S_{\Phi^{\pm}})$ is $|z^3-z\cdot zz|$. Studying the sign
of $z^3-z\cdot zz$, we realize that before the cusp seen in 
Fig. \ref{fig_min_geral2} we have $z^3-z\cdot zz < 0$ while after it we have
$z^3-z\cdot zz > 0$. The change of sign of $z^3-z\cdot zz$ leads to a 
discontinuity in the derivative of $\overline{D}(S_{\Phi^{\pm}})$ with respect to 
$\lambda$ and this gives rise to the cusp
seen in Fig. \ref{fig_min_geral2}. When $\gamma=1.0$, however, we have
$z^3-z\cdot zz \leq 0$ when $0.0\leq \lambda \leq 2.0$ 
and this is why we do not see a similar cusp in Fig. \ref{fig_min_geral3}.

\begin{figure}[!ht]
\includegraphics[width=8cm]{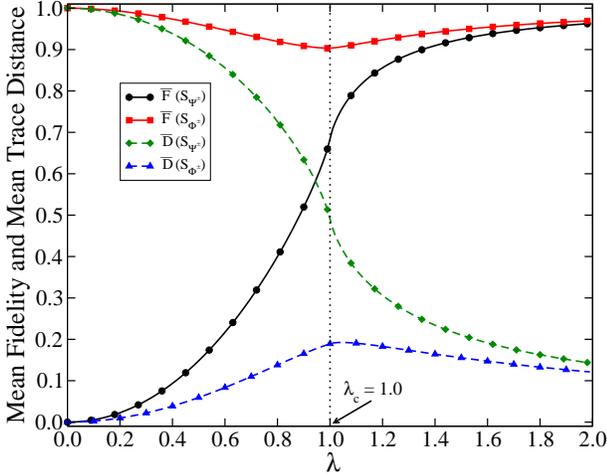}
\caption{\label{fig_extra3}(color online) Same as Fig. \ref{fig_extra2} but
now we have $\gamma=1.0$ (Ising transverse model).}
\end{figure}

Studying the behavior of $z^3-z\cdot zz$ and of $z$, specifically their sign 
for $0.0\leq \lambda \leq 2.0$, we can simplify the expressions for 
$\overline{\mathcal{D}}$ within that range of values for $\lambda$. When 
$\gamma=0$, we have $z^3-z\cdot zz\geq 0$ and $z>0$ \cite{pav23}. Thus, Eq.~(\ref{dmin2}) becomes 
\begin{equation}
\overline{\mathcal{D}}_{\gamma=0} = \left\{ 
\begin{array}{ccc}
z(1-zz)/2, &\mbox{if}& zz \geq -z^2, \\
z(1+z^2)/2, &\mbox{if}& zz \leq -z^2.
\end{array}
\right. 
\end{equation}
Note that the condition $zz=-z^2$ is the solution to 
$\overline{D}(S_{\Phi^{\pm}})=\overline{D}(S_{\Psi^{\pm}})$, which 
determines one of the crossing points shown in Fig. \ref{fig_extra1}.

If we now set $\gamma=0.5$, we have that $\overline{\mathcal{D}}$ is always
given by $\overline{D}(S_{\Phi^{\pm}})$ (cf. Fig. \ref{fig_extra2}). We also
have for $0.0\leq \lambda \leq 2.0$ that $z>0$ \cite{pav23}. 
Noting that in this scenario $z^3-z\cdot zz$ changes sign at $zz=z^2$, 
we obtain for Eq.~(\ref{dmin2}),
\begin{equation}
\overline{\mathcal{D}}_{\gamma=0.5} = \left\{ 
\begin{array}{ccc}
z(1-z^2)/2, &\mbox{if}& zz \geq z^2, \\
z(1-zz)/2, &\mbox{if}& zz \leq z^2.
\end{array}
\right. 
\end{equation}

When $\gamma=1.0$ and $0.0\leq \lambda \leq 2.0$, 
we also have $z>0$ \cite{pav23} and $\overline{\mathcal{D}}$ 
given by $\overline{D}(S_{\Phi^{\pm}})$ (cf. Fig. \ref{fig_extra3}). In this case, though, we always get that
$z^3-z\cdot zz \leq 0$. Therefore, 
\begin{equation}
\overline{\mathcal{D}}_{\gamma=1.0} = z(1-z^2)/2. 
\end{equation}

\subsection{The $\gamma$ transition}

\begin{figure}[!ht]
\includegraphics[width=8cm]{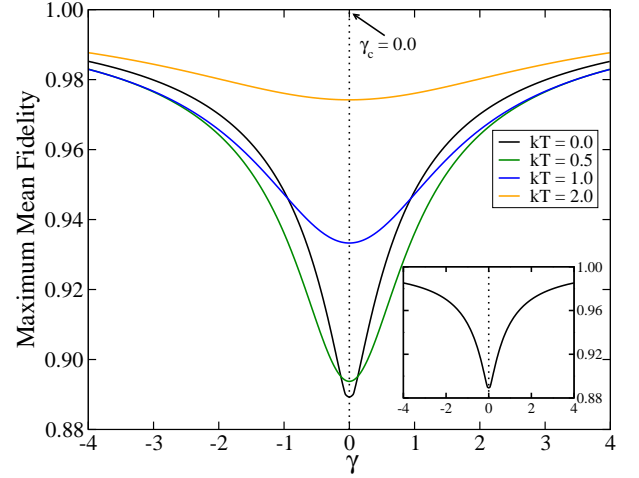}
\caption{\label{fig_max_geral_gamma}(color online) 
$\overline{\mathcal{F}}$, Eq.~(\ref{fmax2}), as a function of
$\gamma$ with $\lambda=1.5$ [see Eq.~(\ref{Hxy})]. 
At $T=0$ (see inset) and $T>0$, the QCP $\gamma_c=0.0$ is detected by a
minimum that occurs at the exact location of the QCP. The dotted lines spotlight 
the QCP $\gamma_c$ and for the solid curves the temperature increases from bottom 
to top along the dashed line.}
\end{figure}

\begin{figure}[!ht]
\includegraphics[width=8cm]{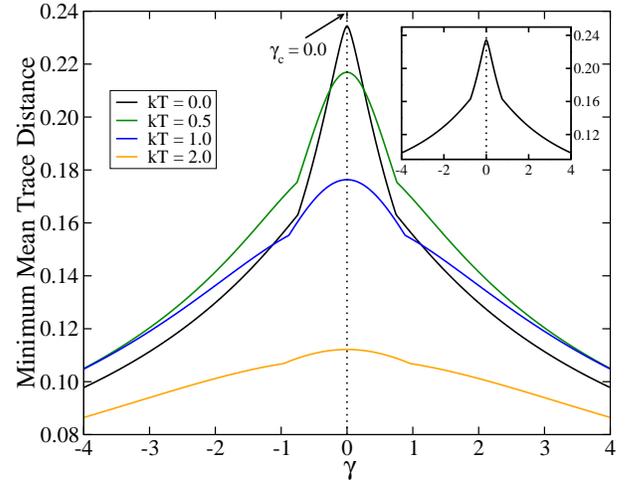}
\caption{\label{fig_min_geral_gamma}(color online) 
Same as Fig. \ref{fig_max_geral_gamma} but now we have 
$\overline{\mathcal{D}}$, Eq.~(\ref{dmin2}),
as a function of $\gamma$. In this case the QCP is detected by a cusp at $T=0$
(see inset), 
with the latter being smoothed out as we increase $kT$. For the solid curves,
the temperature increases from top to bottom along the dashed line.}
\end{figure}

In Figs. \ref{fig_max_geral_gamma} and \ref{fig_min_geral_gamma}
we plot, respectively, $\overline{\mathcal{F}}$ 
and $\overline{\mathcal{D}}$ as functions of $\gamma$ while fixing 
$\lambda=1.5$. 


Looking at Fig. \ref{fig_max_geral_gamma}, we note that the QCP located at
$\gamma_c=0.0$ is given by the global minimum of $\overline{\mathcal{F}}$,
which occur exactly at $\gamma_c$ even for finite $T$. 
If we now look at Fig. \ref{fig_min_geral_gamma}, we realize that 
the global maximum of $\overline{\mathcal{D}}$ marks the location of the QCP.
At $T=0$ we also have a cusp at the maximum, which is smoothed 
out as we increase $T$. However, the maximum of $\overline{\mathcal{D}}$ is not
displaced as we increase $T$, being always situated at the exact location of the 
QCP. 
There are two other cusps in the curve for $\overline{\mathcal{D}}$ that 
are not related to QPTs. They occur when $z^3-z\cdot zz$ changes sign.
Before the first cusp and after the last cusp we have $z^3-z\cdot zz<0$
while between those cusps we have $z^3-z\cdot zz>0$. Since 
$\overline{\mathcal{D}}$ depends on the magnitude of $z^3-z\cdot zz$, these 
sign changes lead to two discontinuities of 
$d\overline{\mathcal{D}}/d\gamma$ as we vary $\gamma$. These 
discontinuities in the derivatives of $\overline{\mathcal{D}}$ manifest themselves
as two cusps in the curves for $\overline{\mathcal{D}}$ that are not related 
to QPTs. 

The robustness of $\overline{\mathcal{D}}$ to spotlight the anisotropy 
transition as we increase $T$ 
can be understood by noting three things that are simultaneously 
true in the neighborhood
of the QCP for all the values of temperature shown in 
Fig. \ref{fig_min_geral_gamma}. By the neighborhood of the QCP 
we mean the region between the two cusps depicted in 
Fig. \ref{fig_min_geral_gamma} that are not related to QPTs.  

First, in the neighborhood of the QCP we have that $\overline{\mathcal{D}}=\overline{D}(S_{\Phi^{\pm}})$. Second, 
$z>0$ and $z^3 - z\cdot zz \geq 0$. Third, 
$z$ has a maximum at $\gamma_c$ while $zz$ has a minimum there \cite{pav23}.  
The first two facts imply that
\begin{eqnarray}
\overline{\mathcal{D}}_{\lambda=1.5} = z(1-zz)/2, &\mbox{if}& zz \leq z^2.
\label{Dlambda}
\end{eqnarray}
The third fact gives $\dot{z}=\dot{zz}=0$, $\ddot{z}<0$, and $\ddot{zz}>0$
at $\gamma_c$, where the single and double dots denote the first and second order derivatives with respect to $\gamma$. A simple calculation using 
Eq.~(\ref{Dlambda}) and the derivatives above give at $\gamma_c$,
\begin{eqnarray}
\frac{d}{d\gamma}\overline{\mathcal{D}}_{\lambda=1.5}(\gamma_c) &=& 0, \\
\frac{d^2}{d\gamma^2}\overline{\mathcal{D}}_{\lambda=1.5}(\gamma_c) &=& 
\frac{\ddot{z}(1-zz)-z\cdot \ddot{zz}}{2} < 0,
\end{eqnarray}
where the last inequality comes from the fact that $\ddot{z}<0$, 
$|zz|\leq 1$, $z>0$, and $\ddot{zz}>0$. In other words, 
$\overline{\mathcal{D}}$ has always a maximum at the QCP for all the values 
of temperature given in Fig. \ref{fig_min_geral_gamma}.

\textit{Remark 8:} A similar analysis explains why the maximum 
mean fidelity $\overline{\mathcal{F}}$ has a minimum at the QCP for all the 
temperatures shown in Fig. \ref{fig_max_geral_gamma}.

\section{Discussion} 
\label{discussion}

The strategy developed in this work to detect quantum critical points (QCPs) using finite temperature data simplifies the teleportation based QCP detectors 
first given in Ref. \cite{pav23}. In Ref. \cite{pav23}, the state to be teleported to one of the qubits of the spin chain was external to it. Here, however, we showed that by using an internal qubit, i.e., a qubit that belongs to the chain itself, we can still detect QCPs with finite $T$ data using the same 
techniques of Ref. \cite{pav23} if we have spin chains embedded in an 
external magnetic field. In this scenario and 
for all spin chain models studied here, we showed that the efficiency of the teleportation protocol or its derivative change drastically at a QCP. 

The present proposal also retains the remarkable features of the former one \cite{pav23}. From a theoretical point of view, all we need to  
calculate the efficiency of the teleportation protocol, quantified by 
either the maximum 
mean fidelity $\overline{\mathcal{F}}$ or the minimum mean trace distance 
$\overline{\mathcal{D}}$, is the state
describing two nearest neighbor qubits of the spin chain [Eq.~(\ref{rhoAB2})],
with the latter assumed to be in equilibrium with a heat bath at temperature $T$. This two-qubit 
density matrix is theoretically computed by tracing out all but these two 
nearest neighbors from the canonical ensemble density matrix describing the  complete chain. 
Equivalently, if we are able to compute all the one- and two-point 
correlation functions we can determine the two-qubit density matrix describing a 
pair of spins since it is completely characterized by these correlation functions. 

The standard approach to study quantum phase transitions also
makes use  of those correlation functions or of quantities that
depend on them (the magnetic susceptibility, for instance). 
The QCPs are detected at $T=0$ by discontinuities in the $n$-th
order derivatives of those quantities at the QCPs. 
Quantum information theory quantities, such as bipartite or
multipartite entanglement and quantum discord, are useful to
spotlight QCPs too
\cite{wu04,oli06,dil08,sar08,fan14,gir14,wer10,wer10b}. 
See also the application of generalized phase-space techniques 
to detect QCPs \cite{mza19,mil23}.
All these quantities are functions of the correlation functions
of the system, the same ones we need to theoretically apply the
present and the method of Ref. \cite{pav23}. 

As already pointed out in Ref. \cite{pav23}, some of the standard quantities used to detect QCPs (magnetization or magnetic susceptibility, for instance) 
become less effective to correctly identify the 
QCPs with finite $T$ data \cite{wer10b}. 
And some quantum information theory quantities, in particular the 
entanglement of formation \cite{woo98}, are zero at and around the QCPs as we 
increase $T$. These quantities cannot be employed to spotlight QCPs
after a certain temperature threshold \cite{wer10b}. On the other hand,
a specific measure of non-classical correlations between two systems, namely, quantum discord \cite{oll01,hen01}, is extremely robust to spotlight QCPs 
at finite $T$ \cite{wer10b}. The present internal  
teleportation based tools to detect QCPs as well as the ones of Ref. \cite{pav23}
share the quantum discord robustness to detect QCPs with 
finite $T$ data.\footnote{Note that it is important to analyze the 
$T = 0$ limit from the finite $T$ data to make sure that no spurious
classical critical point is ``contaminating'' the finite $T$ data.
Only if the finite $T$ data consistently tend to a specific critical 
point as we decrease $T$ can we be reassured that we are properly seeing 
a QCP. See, for instance, Figs. \ref{fig_qcpT}, \ref{fig_qcpTxy1},
\ref{fig_qcpTxy2}, and \ref{fig_qcpTxy3}.}
In addition to that, and contrary to quantum discord, the 
present tools and the ones of Ref. \cite{pav23} do not become computationally intractable if we deal with high spin systems (high dimensional Hilbert spaces)
\cite{pav23}.

Turning our attention to the experimental aspects of the present proposal, 
it is clearly simpler than the external teleportation QCP detector 
tools \cite{pav23}. There is no need to bring a qubit outside from the spin chain
to implement the present proposal. The input state to be teleported belongs to 
the chain itself. 
Indeed, contrary to the external teleportaton
QCP detector, there is no need to repeat the protocol with 
many different input qubits that cover the whole Bloch sphere  \cite{pav23}. For the internal teleportation QCP detector, 
the input state is fixed and belongs to the system under study.
Also, the present scheme and that of Ref. \cite{pav23} have 
a straightforward experimental interpretation since they are 
built on the
quantum teleportation protocol. Quantum discord,
though, has no direct experimental procedure to its determination.  

To execute the internal 
teleportation protocol, and thus the present 
proposal, we need to be able to implement single qubit measurements, Bell state measurements, and 
local unitary operations on the qubits that belong to a spin chain. The single qubit measurement we need is a projective 
measurement onto the eigenstates of the $\sigma^z$ observable
(our computational basis). 
A Bell state measurement is a joint measurement
involving two qubits that aims to project the two qubits onto one of the 
four Bell states. This measurement can be divided into the following sequence 
of three steps \cite{nie00}. First, we implement a controlled-NOT (CNOT) gate on 
the pair of spins we will be measuring, now called control and target qubits. 
The CNOT is a two-qubit (non-local) gate
whose ``truth table'' in the standard basis is such that it flips the spin of the target qubit if, and only if, the control qubit is given by $|1\rangle$.
The standard basis can be, for instance, the eigenstates of the spin operator 
$\sigma^z$. Second, we implement a Hadamard gate on the control qubit of the 
CNOT gate. This is a single (local) qubit gate whose truth table is 
$(|0\rangle +(-) |1\rangle)/\sqrt{2}\rightarrow |0\rangle(|1\rangle)$. 
Third, we measure the two qubits 
in the standard (computational) basis $\{|00\rangle,|01\rangle,|10\rangle,|11\rangle\}$. Each one of the four previous outcomes in the standard basis 
implies that we projected the two qubits 
onto one of the four Bell states \cite{nie00}. The single 
qubit and Bell measurements 
are Alice's job while Bob's job is to measure his qubit after implementing the 
appropriate local operation (local gate) as given by Eqs.~(\ref{s1})-(\ref{s4}).
All these gates, the ones needed by Alice and Bob, 
are already being implemented in spin chain-like systems using 
state of the art techniques \cite{ron15,bra19,xie19,noi22,xue22,mad22,xie22}.
We believe that in the near future they can be implemented in the sequence
described above to execute the teleportation protocol and consequently the present
proposal. 
We should remark that the experiments reported in Refs.
\cite{ron15,bra19,xie19,noi22,xue22,mad22,xie22} were done 
using small systems. However, some of them implemented 
efficiently the one- and two-qubit gates at room temperature. 

Furthermore, using superconducting qubits to realize 
a spin chain, it is already possible to experimentally 
prepare tens of spins simultaneously with coherence times around 
$100 \mu s$. Also, single and two-qubit gates in this platform 
can be implemented within $10 ns$ to $100 ns$, leading to the execution 
of $10^3$ to $10^4$ gates per coherence time \cite{gam17}. 
We can also implement spin chains experimentally using quantum dots and wells. For quantum wells built with GaAs, at room temperature 
the relaxation time is of the order of a few nanoseconds \cite{ohn99}. 
And for GaAs quantum dots, the relaxation time for a qubit is 
about 50 $\mu s$ at $T\approx 20 mK$ \cite{han03} while for arrays of
Ge/Si quantum dots we have a relaxation time of about $10 \mu s$ at 
$T \approx 5K$ \cite{zin10}. These facts, together with the fact that 
in silicon quantum dots at $T \approx 1K$ we can implement single and two-qubit gates in less than
$100 ns$ \cite{pet22}, imply that at low temperatures ($\approx 1 K$)
and for silicon quantum dots it is possible to execute one hundred 
or so gates before the spin chain thermalizes. This number of gates 
is at least one order of magnitude more than what we need to implement  
the present proposal, where only a few gates (fewer than ten) 
are needed at a given run of the teleportation protocol.\footnote{The
quantum circuit representing all the steps of the teleportation protocol
can be seen in Ref. \cite{nie00} while a review on thermalization is
given in Ref. \cite{mor18}. See also Ref. \cite{pek23} for a
quantum circuit realization of a heat bath in solid quantum systems. 
Furthermore, with the advent of a quantum computer, it will be possible to efficiently simulate and even prepare 
thermal states \cite{bra20,sag21,bra23,wan23}, which in turn 
can be used to implement the present QCP detection strategy.}

Note that the time span to execute all the steps of 
the teleportation protocol plus Bob's measurement of the teleported 
qubit should not exceed the ``relaxation time'' of the spin chain. 
That is, all the steps should be performed before the spin chain returns 
to its equilibrium state with the heat bath. This equilibration time 
depends on the system's Hamiltonian, its initial state, and how it 
interacts with the heat bath. Its theoretical computation is very 
challenging, lying beyond the scope of the present work. Experimentally, 
it can be determined by perturbing the system appropriately using the 
above gates and measuring how long it takes to return to its equilibrium 
state.

We should also contrast the present proposal with the standard way 
of studying QPTs from the experimental point of view. In the standard way,
we must measure the one- and two-point correlation functions using, for 
instance, neutron scattering techniques. The non-analytic properties of 
these one- and two-point correlation functions at the QCP mark a QPT. 
In the present proposal, we need only to measure one-point correlation
functions after ``disturbing'' the system in a specific way. Indeed, to 
reconstruct the state describing his qubit after the teleportation protocol
(the ``disturbance'' in the spin chain), Bob only
needs one-point correlation functions. Any single qubit density matrix is completely
characterized once we know all the one-point correlation functions 
$\med{\sigma^x}, \med{\sigma^y}$, and $\med{\sigma^z}$ \cite{nie00}. 
And in all the models investigated here, we only need one correlation function,
$\med{\sigma^z}$ to be specific. This is true since the qubit with Bob after 
the teleportation protocol is diagonal in the standard basis 
[see Eqs.~(\ref{bob1})-(\ref{bob4})]. In other words, we are trading the
measurement of two-point correlation functions to experimentally determine 
a QCP for the measurement of only one 
one-point correlation function after disturbing the system in a very particular 
way, namely, after applying all the gates associated with the internal quantum teleportation protocol. 

Finally, we want to emphasize that the previous paragraph in no
way implies that the present proposal is experimentally 
simpler or easier than the standard approach to detect QCP, namely, the direct measurement of one- and two-point correlation
functions. Actually, the 
internal teleportation based QCP detector can be applied to detect 
a QCP whether or not it is experimentally implemented as described
in the previous paragraphs. The same standard 
experimental 
techniques to measure one- and two-point correlation
functions serve as well. Once we have those correlation 
functions, we automatically obtain Eq.~(\ref{rhoAB2}), from 
which the maximal mean fidelity $\overline{\mathcal{F}}$
and the minimum mean trace distance $\overline{\mathcal{D}}$
associated with the internal teleportation based 
QCP detector can be computed. And if we look at 
Eqs.~(\ref{fmax2}) and (\ref{dmin2}), we realize that it is the 
non-linear dependence of $\overline{\mathcal{F}}$ and 
$\overline{\mathcal{D}}$ on the one- and two-point correlation functions that sets them apart from the external teleportation QCP detector, where the fidelity is a simple linear function of a 
specific two-point correlation function. We believe that this
unique functional form of $\overline{\mathcal{F}}$ and 
$\overline{\mathcal{D}}$ is the reason of why the internal 
teleportation QCP detector outperforms the external one most of 
the time.

\section{Conclusion}
\label{conclusion}

We simplified the teleportation based quantum critical point (QCP) detector
of Ref. \cite{pav23} in a very important way, reducing the technical demands 
to its possible experimental realization. Instead of using an external 
qubit from the spin chain as the input state to be teleported to the chain, 
in this work we employed a qubit within the spin chain itself. 
Several spin-1/2 chain models 
in the thermodynamic limit were used to test the efficacy of this new approach
in detecting a QCP at zero and non-zero temperatures. We showed that
whenever the spin chain is immersed in an external magnetic field, the 
present ``internal'' teleportation based QCP detector retains all the 
remarkable characteristics of the ``external'' teleportation based QCP detector 
of Ref. \cite{pav23}. In particular, we showed that the present strategy detects
the QCPs for the XXZ model in a longitudinal field, the Ising transverse model, 
the isotropic XX model in a transverse field, and the anisotropic XY model in a transverse field.
Note that we worked with these models because their QCPs were previously characterized by other methods. In this way we could
properly compare the predicted location for the QCPs coming from the present tool with the well-known and established locations 
for those QCPs.

The main idea behind the present and the proposal of Ref. \cite{pav23} is to
use a pair of qubits from the spin chain as the entangled resource shared 
between Alice and Bob in the implementation of the teleportation protocol. 
In this work, an internal qubit of the chain, adjacent to the pair of qubits 
shared between Alice and Bob, is teleported to Bob. 
Similarly to what we saw in Ref. \cite{pav23}, we proved that the efficiency 
of the teleportation protocol changes considerably as we cross the QCP.
In other words, the efficiency of the teleportation protocol depends on 
the quantum phase in which the spin chain lies. The efficiency of the teleportation
protocol in this work was quantified by using both the fidelity and the trace 
distance between the input and output states, namely, between Alice's input
state and Bob's final qubit at the end of the teleportation protocol.

For the several models investigated here, we observed that at $T=0$ the maximum 
mean fidelity $\overline{\mathcal{F}}$ and the minimum mean trace distance
$\overline{\mathcal{D}}$ between the teleported and the input states possess 
a cusp or an inflection point exactly at the QCPs. When we increase $T$, these cusps 
are smoothed out and both the cusps and the inflection points are displaced from 
the exact locations of the QCPs. At finite temperatures, these cusps and inflection points manifest themselves in a high value for the magnitudes of the first and second order derivatives of $\overline{\mathcal{F}}$ and 
$\overline{\mathcal{D}}$ around the exact locations of the QCPs. We also 
showed that for small values of temperature, the extrema of these
derivatives lie close together and by taking the zero temperature limit we
are able to correctly predict the $T=0$ location of the QCP.

Furthermore, the internal teleportation based QCP detector shares
the same features of the external one and of the 
quantum discord, when the latter is also 
employed as a tool to detect QCPs \cite{wer10b}. 
These tools can be used without the knowledge of the order parameter 
associated with the quantum phase transition being investigated  
and they are very resilient to the increase of temperature, giving us useful information at finite $T$ that allows us to predict the exact value of the QCP
at $T=0$. 
On top of that, and contrary to the quantum discord, 
the QCP detectors here developed and in Ref. \cite{pav23} 
have a straightforward experimental meaning and are
scalable to high spin systems \cite{pav23}.

We want to end this work by calling attention to 
a feature already highlighted in Ref. \cite{pav23} and that 
is also observed here. This feature is related to the fact that
for each model investigated
in this work we have a distinctive functional form for
$\overline{\mathcal{F}}$ and $\overline{\mathcal{D}}$. For some models we 
have extra cusps for these functions that are not related to a quantum 
phase transition, for other models the QCP is detected via an inflection point
of $\overline{\mathcal{F}}$ and $\overline{\mathcal{D}}$, and for others 
we have the QCP being detected by discontinuities in their derivatives.
Putting it differently, the ``fingerprint'' of a quantum phase, a quantum 
phase transition, and the underlying model dictating these phases and 
transitions are unique. The functional forms of 
$\overline{\mathcal{F}}$ and $\overline{\mathcal{D}}$ as we drive the system 
around its phase space are specific for a given model.
A systematic study $\overline{\mathcal{F}}$ and $\overline{\mathcal{D}}$
enables the proper detection of the QCP using finite $T$ data 
and, interestingly, 
also the discovery of the underlying spin chain model giving rise 
to those quantum phases and quantum phase transitions. 

We should
also remark that the above features are still present for spin
chains away from the thermodynamic limit. Preliminary calculations
using spin chains containing about 10 qubits already show that
we can detect the QCPs with reasonable accuracy and using finite
temperature data only \cite{pav24b}. 
Moreover, in order to properly exclude 
finite size effects or spurious discontinuities in the derivatives of $\overline{\mathcal{F}}$ and $\overline{\mathcal{D}}$
that are not related to QCPs, both the external and internal teleportation based QCP detectors must be used \cite{pav24b}. 
Only discontinuities in the 
derivatives that occur in both quantities at the same place 
are the ones marking a QCP. Spurious discontinuities or finite 
size effects not related to a QCP do not occur at the same spot   in both quantities \cite{pav24b}. In other words,
the internal teleportation QCP detector not only behaves
differently from the external one but it is crucial in the task 
of identifying finite size effects 
or spurious discontinuities in 
the fidelity and in the trace distance that are not associated with QCPs.

\begin{acknowledgments}
GR thanks the Brazilian agency CNPq
(National Council for Scientific and Technological Development) for funding and 
CNPq/FAPERJ (State of Rio de Janeiro Research Foundation) for financial support
through the National Institute of Science and Technology for Quantum Information.
GAPR thanks the São Paulo Research Foundation (FAPESP)
for financial support through the grant 2023/03947-0.

\end{acknowledgments}




\end{document}